\documentclass[conference, compsoc]{IEEEtran}
\usepackage{times}
\usepackage{graphicx}
\usepackage{makeidx}
\usepackage{amsfonts}
\usepackage{algorithm}
\usepackage{algorithmic}
\usepackage{verbatim}

\usepackage{amsmath}
\usepackage{amssymb}
\usepackage{booktabs}
\usepackage{multicol}

\usepackage{ccaption}

\newcommand{\Agent}{{\cal A}}
\newcommand{\Role}{{\cal R}}

\newcommand{\report}{}
\newcommand{\reportint}{The proofs of the propositions and corollaries, and the algorithms
for the transformation of protocols are to be found in the
appendix.}
\newcommand{\reportappendix}{the appendix }

\begin{document}

\title{Enhancing the Security of Protocols against Actor Key  Compromise Problems}

\author{
\IEEEauthorblockN{Jing Ma} \IEEEauthorblockA{SKLCS, Institute of Software \\
Chinese Academy of Sciences \&\\
University of Chinese Academy of Sciences\\
} \and \IEEEauthorblockN{Wenhui Zhang}
\IEEEauthorblockA{SKLCS, Institute of Software \\
Chinese Academy of Sciences\\
Beijing, China} }

 \maketitle
\begin{abstract}
Security of complex systems is an important issue in software
engineering. For complex computer systems involving many actors,
security protocols are often used for the communication of sensitive
data.
Actor key compromise (AKC) denotes a situation where the
long-term secret key of an actor may be known to an adversary for
some reasons. Many protocols are not secure enough for ensuring
security in such a situation. In this paper, we further study this
problem by looking at potential types of attacks, defining their
formal properties and providing solutions to enhance the level of
security. As case studies, we analyze the vulnerabilities (with
respect to potential AKC attacks) of practical protocols, including
PKMv2RSA and Kerberos, and provide solutions to enhance the level of
security of such protocols.
\end{abstract}

\section{Introduction}

Security of complex systems is an important issue in software
engineering. For complex computer systems involving many actors,
security protocols are often used for the communication of sensitive
data. However, security protocols are not always secure enough,
because of reasons including that there may be weakness in the
methods for generation of secrete keys, storage of keys and so on.
If an actor's key is revealed and used by an adversary to
impersonate another party communicating with the actor, then there
is a key compromise impersonation (KCI) attack \cite{01}, and then
the attacker may obtain sensitive data through such an
impersonation.
Actor key compromise (AKC) attack is a generalization of this kind
of attacks.
This has been studied in \cite{02}, where this property is
formalized and conditions under which it can and cannot be achieved
are identified.

Previous works focus on KCI attacks in the domain of key
establishment protocols. In \cite{03} and \cite{04}  some concrete
two-party protocols have been studied  and  countermeasures to
prevent such attacks provided. The type of KCI attacks is classified
in \cite{05} and \cite{06} based on whether the responder
authenticates the initiator, and use digital signatures and
time-stamps as a help. \cite{07} is the first to study security
attribute of group key exchange protocols under KCI attack. The
first computation model of KCI is provided in \cite{09}. Then
\cite{02} provides a systematic analysis of the consequences of
compromising the actor's secret key and countermeasures, and shows
both constructive and impossibility results.

There are additional issues that need to be  investigated. Firstly,
the classification of  KCI attacks  based on adversary's capability
of eavesdropping and sending messages is  generic and may not reveal
the particular feature of such attacks. Furthermore, providing
definitions of attack types may make it easier to analyze the
vulnerabilities and then modify the protocol for enhancing security.
Second, the work in \cite{02} focuses on the problem where a given
actor may have the secret key being compromised, and  we focus on
solutions for enhancing the security in case one of the actors
(however, which one is unknown) has the secret key being
compromised, and we also consider multi-party protocols and a
different type of security claims.
Third, no practical algorithms have been provided in transforming a
protocol into an AKC resilience one, which is also important for the
practical use of the methods.


The purpose of this work is to provide practical solutions for
transforming protocols to achieve higher security levels against AKC
attacks. The work includes classifying types of AKC attacks and
providing their formal definitions, furnishing solutions, and
providing practical algorithms.
%
%
%

The rest of this paper is organized as follows. Section 2 introduces
the modeling framework and gives formalization of security
properties. In Section 3, we classify four types of AKC attacks and
give formal definition of the attacks. In Section 4, we propose
solutions to prevent such attacks. We present case studies in
section 5 and concluding remarks in Section 6.

\reportint

\section{Preliminaries}

We follow the formal framework for protocol specification and the
execution model defined in \cite{10}\cite{11}.

\subsection{Protocol Specification}

A partial function from $X$ to $Y$ is denoted $f:X\hookrightarrow
Y$. The domain and range of $f$ are denoted $dom(f)$ and $ran(f)$,
respectively. $f[a\mapsto b]$ denotes a function $f'$ such that
$f'(a)=b$, and otherwise it coincides with $f$. We write $\langle
s_0,...,s_n\rangle$ to denote the sequence of elements from $s_0$ to
$s_n$.

Let \emph{\Agent}, \emph{\Role}, \emph{Fresh}, \emph{Var},
\emph{Func}, and \emph{TID} denote sets of agents, roles, Fresh and
so on. \emph{TID} contains two distinguished thread identifiers,
\emph{Test} and $\emph{tid}_\emph{A}$ which stands for a thread of
an arbitrary agent and that of an adversary thread.

$\emph{t}^{\sharp\emph{tid}}$ binds the local term $t$ to the
protocol thread identified by $tid$. By $pk(X)$ we denote X's
asymmetric long-term public key, and $sk(X)$ denotes the
corresponding secret key. The superscript $n$ in $Func(Term^n)$
denotes the arity of parameter. $Const$ is a special case of $Func$
with arity $0$. The use of symmetric cryptography and hashing is not
sufficient to ensure AKC resilience \cite{02}. For brevity, we do
not consider symmetric cryptography in this paper and therefore omit
symmetric cryptographic terms in the definition of the basic
elements of protocols.

\par \noindent \textbf{Definition 1 (Terms):}
\begin{align}
Term ::= &  \notag
   \: \Agent \:|
   \: \Role \: |
   \: Fresh \:|
   \: Var \\
&\notag |
   \: Fresh^{\sharp TID} \:|
   \: Var^{\sharp TID}\: | \: Func(Term^n) \\
   &\notag  |
   \:(Term, Term) \: | \:\{Term\}_{Term} \\
& \notag | \: sk(\Agent) \:| \: pk(\Agent) \:|\: sk(\Role) \:| \:
pk(\Role)
\end{align}

We define $RoleTerm$ as the set of terms that have no subterms in
$\Agent \cup Fresh^{\sharp TID}$, and $RunTerm$ as the set of terms
that have no subterms in $\Role \cup Fresh$. A role term is
transformed into a run term by applying an instantiation from the
set $Inst$: $$TID \hookrightarrow ((\Role\hookrightarrow \Agent)\cup
((Fresh\cup Var)\hookrightarrow RunTerm)).$$

We define a binary relation $\vdash$ on terms, where $M\vdash t$
denotes that the term $t$ can be inferred from the set of terms M.
Let $t^{-1}$ denote the inverse function on terms such that for all
agents $a$, $(pk(a))^{-1}=sk(a)$ and $(sk(a))^{-1}=pk(a)$, and for
all other terms, $t^{-1}=t$. Let $t_0,...,t_n\in Term$ and let $f\in
Func$. The relation $\vdash$ is the smallest relation satisfying:
\begin{align}
\notag t\in M\Rightarrow M\vdash t \\
\notag
M\vdash t_1\wedge M\vdash t_2 \Leftrightarrow M \in (t_1,t_2)\\
\notag M\vdash \{t_1\}_{t_2}\wedge M\vdash (t_2)^{-1}\Rightarrow M\vdash t_1 \\
\notag M\vdash t_1\wedge M\vdash t_2 \Rightarrow M\vdash \{t_1\}_{t_2}\\
\notag \bigwedge_{0\leq i \leq n} M\vdash t_i \Rightarrow M\vdash
f(t_0,...,t_n)
\end{align}

The subterm relation $\sqsubseteq$ is defined as the reflexive,
transitive closure of the smallest relation satisfying the
following, for all terms $t_1,...,t_n$ and function names $f$:
\begin{align}
\notag &t_1\sqsubseteq (t_1,t_2),\: t_2\sqsubseteq (t_1,t_2)\\
\notag & t_1\sqsubseteq\{t_1\}_{t_2},\:
t_2\sqsubseteq\{t_1\}_{t_2} \\
\notag &t_1\sqsubseteq pk(t_1),\: t_1\sqsubseteq sk(t_1) \\
\notag & t_i\sqsubseteq f(t_1,...,t_n)\: for\: 1\leq i \leq n
\end{align}

The accessible subterm relation $\sqsubseteq_{acc}$ identifies
potentially retrievable subterms, is defined as a subset of subterm
relation such that $t_1\sqsubseteq_{acc}(t_1,t_2)$ and
$t_2\sqsubseteq_{acc}(t_1,t_2)$. In order to identifies position of
$pk(a)$ and $sk(a)$, we define another subterm relation
$\sqsubseteq_{ace}$ such that $t_1\sqsubseteq_{ace}\{t_2\}_{t_1}$.


\par \noindent \textbf{Definition 2 (Event):} Let $Claim$ be a given set of claims including the following claims
$commit$, $running$, $secret$, $nisynch$. Let $Label$ be a set  of
labels. The set of events is defined as follows.

\begin{align}
\notag RoleEvent ::&=send_{Label}(\Role, \Role, RoleTerm) \\
\notag &| \quad revc_{Label}(\Role, \Role, RoleTerm)\quad \\
\notag &| \quad claim_{Label}(\Role,Claim[,\Role][,RoleTerm])\\
\notag RunEvent ::&=create(\Role,\Agent) \\
\notag &| \quad  send_{Label}(\Agent, \Agent, RunTerm)\\
\notag &| \quad revc_{Label}(\Agent, \Agent, RunTerm)\\
\notag &| \quad claim_{Label}(\Agent,Claim[,\Agent][,RunTerm])\\
\notag AdvEvent::&= LKR(\Agent)\\
\notag Event::&=RoleEvent \: | \: RunEvent \: | \: AdvEvent
\end{align}

$RunEvent$ describes how agents start threads, send and receive
messages. $LKR(a)$ is an event where the adversary compromises $a$'s
long term secret key. The $AdvEvent$ is executed in the single
adversary thread $tid_A$.

As an example, the event
$$send_{l}(Alice, Bob, \{n^{\#tid}\}_{sk(Alice)})$$ denotes that
$Alice$ sends $Bob$ a nonce $n^{\#tid}$ in the run $tid$ and
encrypted with its secret key.

 An event $e$ has an event-type and a label which are
denoted $evtype(e)$ and $label(e)$, and the contents of a send-event
$e$ is denoted $cont(e)$.

In order to simplify the typing constraint, in the following, $e,e'$
stand for events,  $\rho,\rho'$ stand for sequence of events, $r,r'$
stand for roles, $a,b$ stand for agents, $l,l'$ stand for labels,
$t,t'$ stand for role terms and run terms (should be clear from the
context), $m,n$ stand for run terms that are used in a message,
$tid,tid_1,tid_2$ for TID. Let $X$ be a set. A sequence $y$ of
elements of $X$ is denoted $y \in X^*$. An element $a$ in a sequence
$y$ is denoted $a\in y$. The operation $\cdot$ denotes the
concatenation of two sequences. The powerset of $X$ is denoted
$pow(X)$.

A sequence of RoleEvent is well-formed, if all variables initialized
in an accessible position in a $recv$ event are not used before that
event. Let $vars(X)$ denote the set of variables appearing in $X$.

\begin{align}
\notag wellformed(\rho)\\
\notag \Leftrightarrow \\
\notag \forall \rho',l,a,b,t,\rho'',v:\\
\notag \rho=\rho '\cdot \langle recv_l (a,b,t)\rangle\cdot \rho ''\\
\notag  \Rightarrow (v\sqsubseteq_{acc}t \Rightarrow v\notin
vars(\rho').
\end{align}

A protocol is a partial function from $\Role$ to $Event^*$ together
with a function that formalizes which terms may be stored in a given
variable. For each role, the sequence of events must be wellformed.

\par \noindent \textbf{Definition 3 (Protocol):}
Let $\Pi:\Role\hookrightarrow Event^*$ and
$type_{\Pi}:Var\rightarrow pow(RunTerm)$. If for all $r\in
dom(\Pi)$, $\Pi(r)$ is wellformed, then $(\Pi,type_{\Pi})$ is a
protocol.

For convenience, we extend the domain of $type_{\Pi}$ to RunTerm
such that $type_{\Pi}(t)$ for a run term $t$ is the set of run terms
such that variables in $t$ is substituted according to the initial
$type_{\Pi}$.

In a protocol, a label $l$ is associated with a send-role and a
receive-role, denoted respectively $sl(l)$ and $rl(l)$, defined by
$sl(l)=r$, if $label(e)=l$ and, $e=send_l(r,r',t) \in \Pi(r)$ or
$e=recv_l(r,r',t)\in\Pi(r')$ for some $t$; $rl(l)=r'$, if $label(e)=l$
and, $e=send_l(r,r',t) \in \Pi(r)$ or $e=recv_l(r,r',t) \in
\Pi(r')$.

\subsection{Execution Model}

Protocol execution is modeled as a labeled transition system
$(State,RunEvent,\rightarrow,s_0)$. A state
$s=(tr_s,AK_s,th_s,\sigma_s)$ consists of a trace $tr_s\in (TID
\times (RunEvent \;\cup\; AdvEvent))$, the adversary's knowledge
$AK_s$, a partial function $th_s\in TID \hookrightarrow RunEvent^*$
and a role and variable instantiation $\sigma_s\in Inst$. We denote
$\sigma_s(tid)$ as $\sigma_{s,tid}$, and $tr_s(i)$ as $tr_{s,i}$
which is the $i$-th event of the trace. The initial state $s_0$ is
$(\emptyset,AK_0,\emptyset,\emptyset)$ where $AK_0=\{a,pk(a)\:|\:
a\in \Agent\}\cup \{n^{\# tid_A}:n\in Fresh\}$ is the initial
adversary knowledge.

The operational semantics of a protocol is defined by a transition
system which are composed of execution rules from Fig 1 with a
selected subset of adversary rules in Fig 2. The $create$ rule
starts a new thread of a protocol role $R$. The $send$ rule sends a
message $m$ to the network and add it to adversary knowledge. The
$receive$ rule accepts message if it match the pattern $pt$. The
$claim$ rule states a security property that is expected to hold.
The $LKR_{actor}$ rule allows the adversary to learn the long-term
keys of the agent executing the test run. %

\begin{figure*}
\centering

\begin{center}
\par $\frac {R \in dom(\Pi)\quad tid \not\in(dom(th))\cup \{tid_A,Test\}\quad \sigma' : \Role\rightarrow \Agent}{(tr,AK,th,\sigma)\rightarrow (tr\cdot \langle (tid,create(R,\sigma(R)))\rangle,AK,th[tid\mapsto \sigma'(\Pi(R))^{\#tid}],\sigma[tid\mapsto \sigma'])}$[$create_\Pi$]
\vspace{0.1cm}
\par $\frac {th(tid)=\langle send_l(a,b,m)\rangle\cdot seq}{(tr,AK,th,\sigma)\rightarrow (tr\cdot \langle (tid,send_l(a,b,m))\rangle,AK\cup\{m\},th[tid\mapsto seq],\sigma)}$[send]
\vspace{0.1cm}
\par $\frac {th(tid)=\langle recv_l(a,b,pt)\rangle\cdot seq\quad dom(\sigma')=vars(pt)\quad (\forall x\in dom(\sigma'))(\sigma'(x)\in type_\Pi(x))\quad AK\vdash \sigma'(pt)}{(tr,AK,th,\sigma)\rightarrow (tr\cdot \langle(tid,recv_l(a,b,\sigma'(pt)))\rangle,AK,th[tid\mapsto \sigma'(seq)],\sigma[tid\mapsto \sigma_{tid}\cup \sigma'])}$[$recv_{type_\Pi}$]
\vspace{0.1cm}
\par $\frac {th(tid)=\langle e \rangle\cdot seq\quad evtype(e)=claim}{(tr,AK,th,\sigma)\rightarrow (tr\cdot \langle(tid,e)\rangle,AK,th[tid\mapsto seq],\sigma)}$[claim]
\vspace{0.1cm}
\par Fig.1. Execution-model rules
\vspace{0.1cm}
\par $\frac {a=\sigma_{Test}(R)\quad a\notin \{\sigma_{Test}(R'):R'\in dom(\Pi)\setminus\{R\}\}}{(tr,AK,th,\sigma)\rightarrow (tr\cdot \langle(tid_A,LKR(a)),AK\cup LTK(a),th,\sigma\rangle)}$$[LKR_{actor\Pi,R}]$
\vspace{0.1cm}
\par $\frac {a\notin \{\sigma_{Test}(R):R\in dom(\Pi)\}}{(tr,AK,th,\sigma)\rightarrow (tr\cdot\langle(tid_A,LKR(a))\rangle,AK\cup LTK(a),th,\sigma)}$$[LKR_{others\Pi}]$
\vspace{0.1cm}
\par Fig.2. Adversary-compromise rules
\end{center}
\end{figure*}

Let the protocol $(\Pi, type_{\Pi})$ with an initial role $R\in
dom(\Pi)$, and a set of adversary rules $A$ be given. If there is a
rule such that $s\rightarrow s'$, then we write
$s\rightarrow_{\Pi,type_{\Pi},R,A}s'$. The set of reachable states
denoted $RS(\Pi,type_{\Pi},R, A)$ is $\{s \:|
\:s_0\rightarrow_{\Pi,type_{\Pi},R,A}^*s\}$. The set of all possible
traces of the protocol $(\Pi, type_{\Pi})$ is denoted $Traces(\Pi,
type_{\Pi})$.

In a state $s$, we have a trace $tr_s$ and each thread in the trace
is created by a role. The special thread $Test$ is created by $R$.
Let $role_s:TID \rightarrow \Role$ be a function that identifies a
$tid$ with a $role$ in $s$. Then  $role_s(Test)=R$ and
$role_s(tid)=r'$, if $(tid, create(r',\sigma_s(r')))\in tr_s$.

\subsection{Security Property}
\par \noindent
Security properties are modeled as reachability properties. A
$secrecy$ claim on a role term $t$ is of the form
$claim_l(r,secret,t)$ for some label $l$ and role $r$.

\par \noindent \textbf{Definition 4 (secrecy claim):}
Let $s$ be a state. If $\gamma= claim_l(r,secret,t)$ is a secrecy
claim on $t$, and $(Test, \sigma_{s,Test}(\gamma^{\#Test}))\in
tr_s$, then
\begin{align}
\notag s \models \gamma \Leftrightarrow AK_s\nvdash
\sigma_{s,Test}(t^{\#Test})
\end{align}

The following two properties are related to data agreement.

The $commit$ property means that the initiator agree on some data
with the responder. The $nisynch$ property means whenever initiator
$I$ completes a run of the protocol with responder $R$, then $R$ has
previously been running the protocol with $I$, and the two agents
agreed on all the variables. A commit claim on a role term $t$ is of
the form $claim_l(r,commit,r',t)$ for some label $l$ and roles $r$
and $r'$.
A corresponding $running$ claim for such a $commit$ claim is of the
form $claim_{l}(r',running,r,t)$.

\par \noindent \textbf{Definition 5 (commit claim):}
Let $s$ be a state. If $\gamma= claim_l(r,commit,r',t)$ is a commit
claim, and $(Test, \sigma_{s,Test}(\gamma^{\#Test}))\in tr_s$, then
$s \models \gamma$, iff
\begin{itemize}
\item there is a $tid$ such that $role_s(tid) = r'$, and
\item there is a  running claim
$\delta= claim_{l}(r',running,r,t)$ such that $(tid, \sigma_{s,Test}(\delta))\in tr_s$, and there
exists a send-event $e$, such that $(tid,e)\in tr_{s}$,
$t\sqsubseteq_{acc}cont(e)$.

\end{itemize}

Let $<_R$ denote the total order of events in a sequence (for the
sequence of events
$\langle\varepsilon_1,\varepsilon_2,\varepsilon_3\rangle$, we have
$\varepsilon_1<_r \varepsilon_2$, $\varepsilon_2<_r \varepsilon_3$,
and $\varepsilon_1<_r \varepsilon_3$). The order on events which is
induced by the communications is defined as
$\varepsilon_1\dashrightarrow \varepsilon_2\Leftrightarrow\exists
l,r,r',t_1,t_2:\varepsilon_1=send_l(r,r',t_1)\wedge\varepsilon_2=recv_l(r,r',t_2)$.
The transitive closure of the union of the role event order and the
communication relation is called the protocol order
$\prec_P\:=(\dashrightarrow\cup\bigcup_{r\in \Role}<_r)^+$.
$prec(cl)$ is the set of causally preceding communications of a
claim event labeled with $cl$:
$prec(cl)=\{l\:|\:recv_l(\_,\_,\_)\prec_p claim_{cl}(...)\}$.

Let $tidinst_s: \Role \hookrightarrow pow(TID)$ denote the function
that maps roles to runs according to $tr_s$ of the state $s$. Let
$ev(tr_{s,i})$ denote $e$ iff $tr_{s,i}=(tid,e)$ for some $tid$.

A $nisynch$ claim is of the form $claim_{l}(r,nisynch)$ for some $r$
and $l$ for stating the correspondence between send-messages and
recv-messages.

\par \noindent \textbf{Definition 6 (nisynch claim):}
Let $s$ be a state. If $\gamma= claim_{l}(r,nisynch)$ is a nisynch
claim, and $(Test, \sigma_{s,Test}(\gamma^{\#Test}))= tr_{s,k}$ for
some $k$, then
\begin{align}
\notag s \models \gamma \Leftrightarrow & \forall l'\in prec(l),\;a,\;b,\;m,\; \\
\notag & \forall j<k,tid\in
tidinst_s(rl(l')):\\
\notag &(ev(tr_{s,j})=recv_{l'}(a,b,m)^{\#tid} \\
\notag & \quad\Rightarrow \exists i<j,tid'\in
tidinst_s(sr(l')):\\
\notag & \quad\quad\quad (ev(tr_{s,i})=send_{l'}(a,b,m)^{ \#tid'})
\end{align}

A protocol is AKC secure if its security claim holds under AKC
attacks. This property has been formalised in \cite{02}. Here we use
$(\Pi, type_{\Pi})\models_A \gamma$ to denote that for all $s\in
RS(\Pi, type_{\Pi}, R, A)$, $s\models \gamma$.

\par \noindent \textbf{Definition 7 (Actor key compromise security, AKCS):}
Let $(\Pi, type_{\Pi})$ be a protocol, $R\in dom(\Pi)$, $A$ an
adversary (represented by a set of adversary rules) such that
$LKR_{actor\Pi,R}\in A$, and $\gamma \in \Pi(R)$ a security claim.
$\gamma$ is $actor$ $key$ $compromise$ $secure$ (AKCS) $in$ $(\Pi,
type_{\Pi})$ $with$ $respect$ $to$ $A$ if $(\Pi,
type_{\Pi})\models_A \gamma$.

For the correctness of security properties, we assume that no
asymmetric long-term secret keys appear in accessible positions in
any messages of a protocol, in the subsequent sections.

\section{Attack Types}

Understanding adversary's techniques to launch attacks and their
attack objectives is helpful in identifying weakness of protocols.
Some work has been done on categorizing attacks with traditional
Dolev-Yao adversary model. In \cite{12}, there is a classification of
known-key attacks, where they study AK protocols and categorize
attacks based on adversary's capability of modifying messages. In
\cite{13}  one-pass two-party key establishment protocols under KCI
attacks are studied, two classes of KCI attacks are described. Here
we study types of attacks under stronger adversary models.
Furthermore, we provide the formal definition of  such attacks based
on the trace model and techniques for fixing such protocols are
provided in the next section.

\par \noindent \textbf{Secrecy Attack}
One purpose of a protocol is to transmit a secret nonce from an
initiator to a responder. In order to keep the nonce secret, The
initiator will encrypt the nonce with the responder's public key,
which is not safe if intruders knows the responder's secret keys.

\par \noindent \textbf{Definition 8 (Secrecy attack):}
\par \noindent Let $(\Pi, type_{\Pi})$ be a protocol, $R\in \Role$, $t\in
RoleTerm$. If $\exists s\in RS(\Pi, type_{\Pi},R,A)$, $AK_s \vdash
\sigma_{s,Test}(t^{\#Test})$,  then there is secrecy attack on t,
which we denote $SecrecyAttack(t,\Pi,type_{\Pi})$.

\par \noindent \textbf{Example}
Suppose that the initiator wants to transmit a secret nonce to the
responder before setting up a session key. In order to keep the
nonce secret, the initiator  will encrypt the nonce with the
responder's public key, which is not safe if intruders knows the
responder's secret keys. Consider the CCIT-ban1\cite{19} protocol as follows.
\begin{align}
& \notag I \rightarrow R : \\
& \notag \quad\quad I,
\{Ta,Na,R,Xa,\{Ya,\{hash(Ya)\}_{sk(I)}\}_{pk(R)}\}_{sk(I)}
\end{align}

Clearly, there is secrecy attack on  Ya, if the secret key of the
responder is known to the intruder.

\par \noindent \textbf{Substitution Attack}
An attack of this type occurs in a situation when an initiator and a
responder try to use fresh values or secret keys to authenticate
each other. The main characteristics of this type of attacks is that
the adversary replaces terms in a message with another terms without
being discovered.

Let $Match_s(a,tid_1,b,tid_2)$ denote that the thread $tid_1$
instantiated by the agent $a$ is the corresponding thread
communicating with $tid_2$ instantiated by $b$ according to
$\sigma_s$ of the state $s$. In other words,
$Match_s(a,tid_1,b,tid_2)$ iff there is $r,r'$ such that
$\sigma_{s,tid_i}(r)=a$ and $\sigma_{s,tid_i}(r')=b$ for $i=1,2$.

Let $m[x/y]$ denote $m'$ derived from $m$ by replacing $y$ in $m$
with $x$. Let $L$ be a subset of labels, $S$ and $S'$ be sets of
terms, and $\prec$ be an access relation. The predicate $Replace$ is
defined as follows.
\begin{align}
\notag
Replace(s, L, S, S',\prec,tid)\Leftrightarrow \\
\notag  \exists l\in L,\;m,\;m',\;a,\;b',\;tid',\\
\notag
x\in S,\;y\in S',\; y\not=x,\;x\prec m':\\
\notag  tid' \in tidinst_s(sr(l)) \land tid \in tidinst_s(rl(l)) \land\\
\notag Match_s(a,tid',b,tid)\land\\
\notag
\exists k.(ev(tr_{s,k})=recv_l(a,b,m)^{\#tid}) \land\\
\notag\forall j<k.( ev(tr_{s,j}) =
send_l(a,b,m')^{\#tid'}\\
\notag \Rightarrow m'=m[x/y])
\end{align}

In a substitution attack, the adversary eavesdrop the message and
modify some of its fresh values by its own fresh values and transmit
it to the receiver of the message.

Let $Finish(s,tid)$ denote the thread $tid$ has been completed in
$s$, i.e., every event in the sequence $th_s(tid)$ has a
corresponding event in $tr_s$.

Let $S= \{t \cup f(t) \:|\: t \in Fresh^*,f\in Func\}$ and $S'=\{t
\cup f(t)\:|\: t \in AdvFresh^*,f\in Func\}$, where $AdvFresh$
denote the subset of $Fresh$ used by the adversary.

\par \noindent \textbf{Definition 9 (Substitution Attack):}
For a security protocol $(\Pi, type_{\Pi})$, there is a substitution
attack, if $\exists s\in RS(\Pi, type_{\Pi},R,A)$ and a $tid$ such
that $Finish(s,tid)$ and $Replace(s,Label,S,S',
\sqsubseteq_{acc},tid)$ hold, which we denote $SubAttack(\Pi,
type_{\Pi})$.

\par \noindent \textbf{Example}
Consider the Bilateral Key Exchange (BKE) protocol as an example,
which is supposed to guarantee the secrecy of $kir$ and agreement on
$nr$ and $ni$.
\begin{align}
& \notag 1. \; I \rightarrow R : \{ni,I\}_{pk(R)} \\
& \notag 2. \; R \rightarrow I : \{hash(ni),nr,R,kir\}_{pk(I)}\\
& \notag 3. \; I \rightarrow R : \{hash(nr)\}_{kir}
\end{align}
\par \noindent The protocol is vulnerable to substitution attacks. If the
intruder (denoted $D_{Alice}$) knows the secret key of Bob (an agent
of the role $R$), he can decrypt message 2 using the secret key, and
constructing another message 2' using its own nonces. In this way,
the adversary impersonate Bob to Alice (an agent of $I$) and break
agreement of $ni$ and $nr$ between them:
\begin{align}
& \notag 1. \; Alice \rightarrow Bob : \{ni,Bob\}_{pk(Bob)} \\
& \notag 2. \; Bob \rightarrow D_{Alice} : \{hash(ni),nr,Alice,kir\}_{pk(Alice)}\\
& \notag 3. \; D_{Alice} \text{ decrypts\; message\; using }\;sk(Alice) \;\\
& \notag \quad \text{and \;learns} \;hash(ni)\\
& \notag 4. \; D_{Alice} \text{ chooses}\; nr',\;kir'\; \\
& \notag \quad \text{and\; constructs}\; \{hash(ni),nr',Alice,kir'\}_{pk(Alice)}\\
& \notag 5. \; D_{Alice} \rightarrow Alice : \{hash(ni),nr',Alice,kir'\}_{pk(Alice)}\\
& \notag 6. \; Alice \rightarrow D_{Alice} : \{hash(nr')\}_{kir'}
\end{align}

\bigskip

\par \noindent \textbf{Role-mixup Attack}
An attack of this type  has the result that the participating
entities do not agree on who is playing what role in the protocol.
We use $Termin(s,L)$ to denote that there exists some label $l \in
L$ which contains role name in accessible position and there is no
matching send-events for a recv-event in the trace.
\begin{align}
\notag Termin(s,L,tid)\Leftrightarrow \\
\notag \exists l\in L,\; a,b,\;m,n,\;tid':\\
\notag tid' \in tidinst_s(sr(l)) \land tid \in tidinst_s(rl(l)) \land\\
\notag  Match_s(a,tid', b,tid) \land \\
\notag \exists k.(ev(tr_{s,k})=recv_l(a,b,m)^{\#tid'}) \land \\
\notag \forall j<k,l'\in Label.(ev(tr_{s,j}) =
send_{l'}(a,b,n)^{\#tid}\\
\notag \Rightarrow l\not=l')
\end{align}

The role-mixup attack states that the messages which has agent names
in accessible position have been replaced by the adversary, or the
public(secret) key of some agent may be replace by other agent's
public(secret) key, or the adversary forged a message with agent
names in accessible position to impersonate another party.

\par \noindent \textbf{Definition 10 (Role-mixup attack):}
Let $(\Pi, type_{\Pi})$ be a protocol,  $L$ be the subset of $Label$
such that agent names are accessible in the corresponding events,
i.e. $L=\{l\:|\: \exists a, e.(label(e) = l \land a\sqsubseteq_{acc}
cont(e))\}$, $S = \{pk(a) \cup sk(a) \:|\: a\in \Agent\}$. The
role-mixup attack of $(\Pi, type_{\Pi})$, denoted
$RoleMixupAttack(\Pi, type_{\Pi})$, is defined as follows.
\begin{align}
\notag RoleMixupAttack(\Pi, type_{\Pi}) \Leftrightarrow\\
\notag \exists s\in RS(\Pi, type_{\Pi},R,A),tid: \\
\notag  Finish(s,tid) \land \\
\notag (Replace(s,L,\Agent,\Agent,\sqsubseteq_{acc},tid) \vee  Termin(s,L,tid) \vee \\
\notag
 Replace(s,Label,S,S,\sqsubseteq_{ace},tid))
\end{align}

\par \noindent \textbf{Example}
Consider the isoiec-9798-3-5 \cite{20} protocol as an example:
\begin{align}
& \notag 1. \; A \rightarrow B : Cert(A), RA, Text1\\
& \notag 2. \; B \rightarrow A : Cert(B), RB, Text2\\
& \notag 3. \; B \rightarrow A : RB, RA, A, Text6,\{RB, TA, A, Text5\}_{sk(B)}\\
& \notag 4. \; A \rightarrow B : RA, RB, B, Text4, \{RA, RB, B,
Text3\}_{sk(A)}
\end{align}

\par \noindent
The protocol is vulnerable to role-mixup attacks. In this protocol
Bob and Alice want to agree on fresh values $RA$, $RB$, $Text3$ and
$Text5$.
The attack is shown in Fig 3, in which the adversary listens to the
message between them and impersonate Alice and Bob, such that Alice
assumes Bob as B and Bob assumes Alice as B, however both Alice and
Bob are acting as A.

\begin{figure*}
\centering
\begin{center}
  \captionwidth{0.9\textwidth}
  \changecaptionwidth
  \centering
  \includegraphics[width=1.0\textwidth]{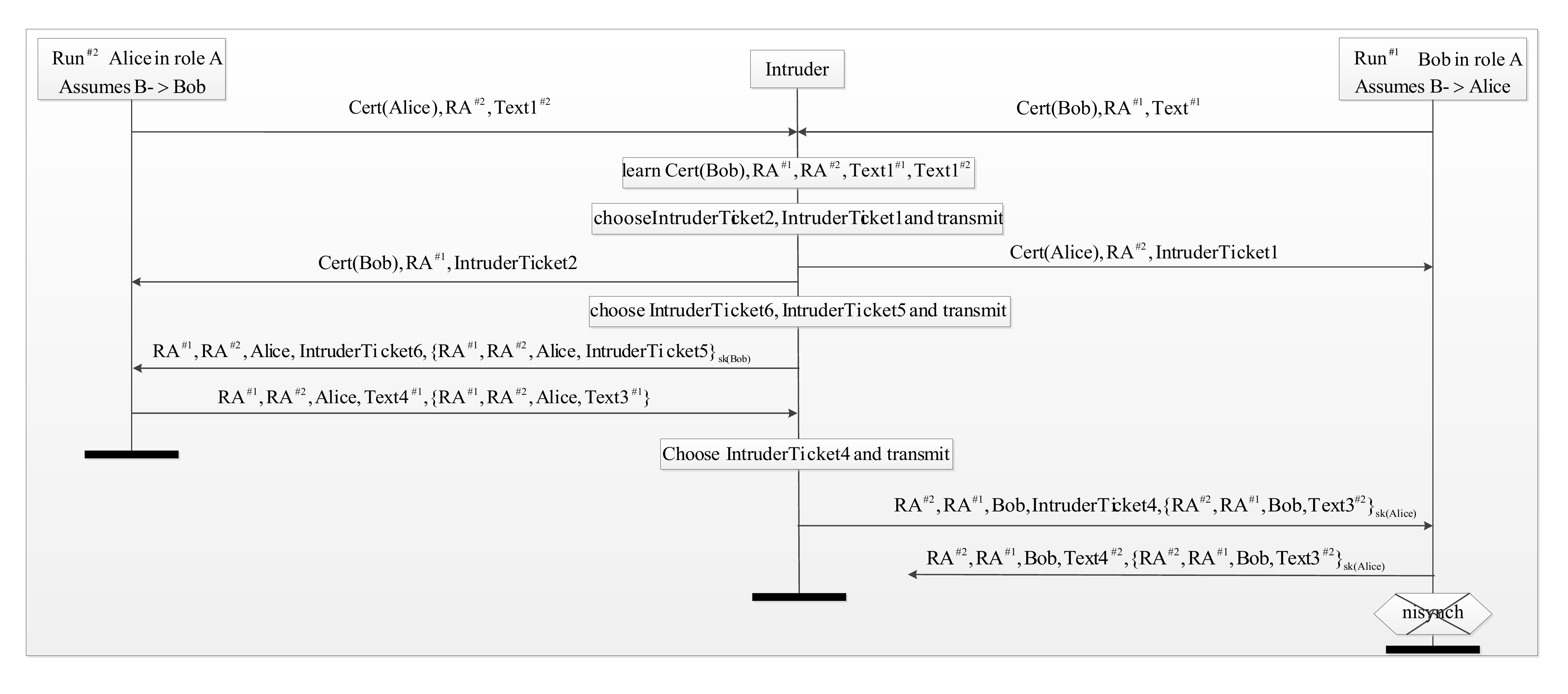}
  \\ {Fig 3}
\end{center}
\end{figure*}

\par \noindent \textbf{Parallel Attack}
In the environment that the same protocol has run as several
threads, the authentication may not be preserved because A may
communicate with B in the first thread, and with C which has run the
same protocol later, but A still assumes he is communicating with B.

\par \noindent \textbf{Definition 11 (Parallel Attack):}
Let $(\Pi, type_{\Pi})$ be a protocol. The parallel attack of $(\Pi,
type_{\Pi})$, denoted $ParallelAttack(\Pi, type_{\Pi})$, is defined
as follows.
%
\begin{align}
\notag
ParallelAttack(\Pi, type_{\Pi})\Leftrightarrow\\
\notag \exists s\in RS(\Pi, type_{\Pi},R,A),\; l,\; a,b,\;m,\; \\
\notag  \exists k,tid\in TID.(ev(tr_{s,k}) =
recv_l(a,b,m)^{\#tid})\wedge\\
\notag \forall j<k,tid'\in tidinst_s(sr(l)):\\
\notag \quad\quad (ev(tr_{s,j}) = send_l(a,b,m)^{\#tid'} \\
\notag \Rightarrow !Match_s(a,tid',b,tid))
\end{align}

\par \noindent \textbf{Example}
Consider the following protocol, in which the two agents
authenticate each other using three nonces.
\begin{align}
& \notag 1. \quad A \rightarrow B : \{na\}_{sk(A)} \\
& \notag 2. \quad B \rightarrow A : \{h(na,nb),nb\}_{sk(B)}\\
& \notag 3. \quad A \rightarrow B : \{h(nb,nc),nc\}_{sk(A)}\\
& \notag 4. \quad B \rightarrow A : \{h(nc)\}_{sk(B)}
\end{align}

The protocol is vulnerable to parallel attack when Alice has two
runs of the protocol. The adversary can forge the message in the
second run, which makes Bob initiate a session with Alice in run 1
but receive the last authentication message in run 2. We show the
attack in Fig 4.
\begin{figure}[ht]
  \captionwidth{0.9\textwidth}
  \changecaptionwidth
  \centering
  \includegraphics[width=0.5\textwidth]{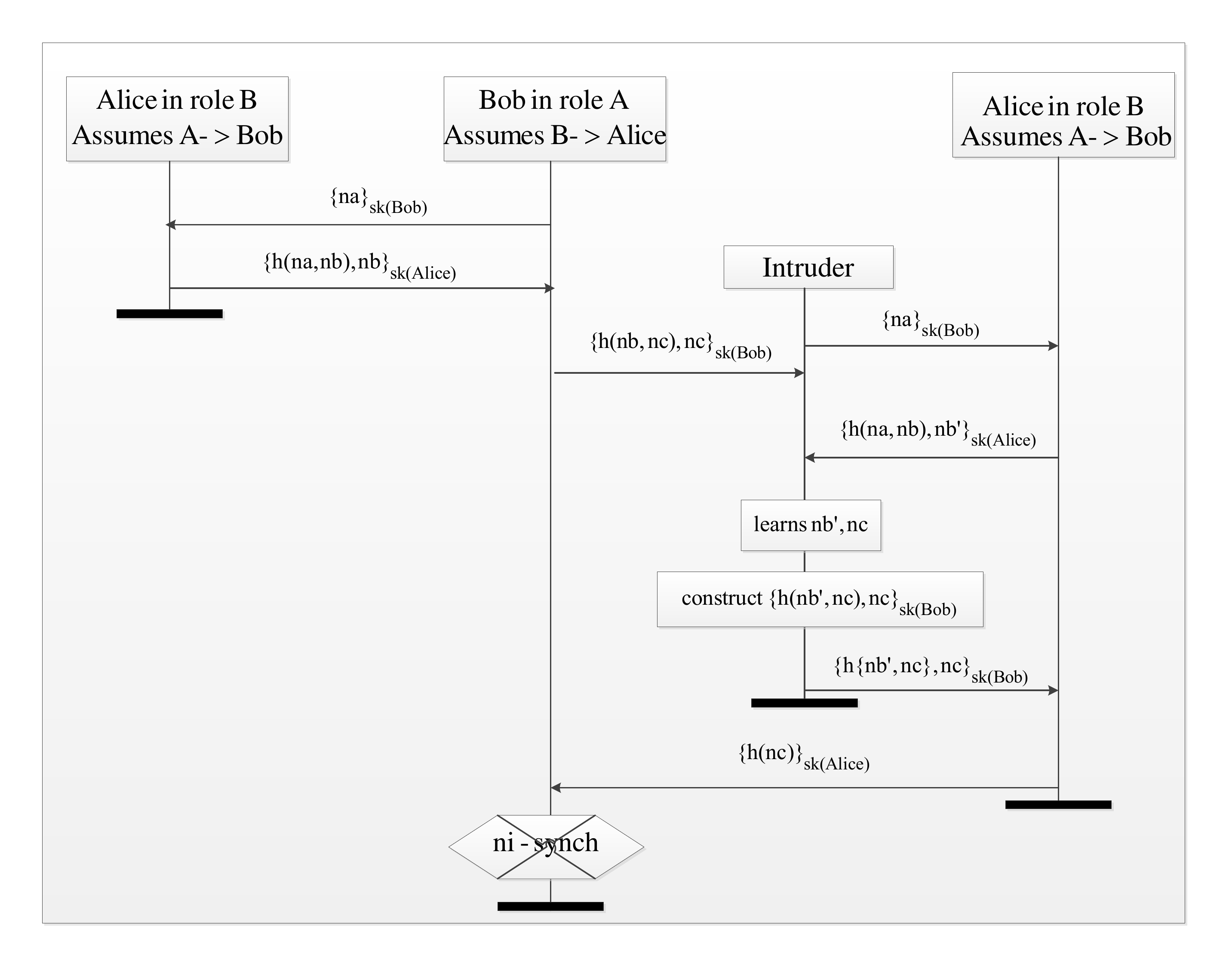}
  \\ {Fig 4}
\end{figure}

\section{Preventing Attacks}

In this section, we give constructive methods for avoiding potential
AKC attacks.
In \cite{02}, transformations to achieve unilateral security is
provided.  Our work tries to provide transformations that achieve
bilateral secrecy and agreement, and instead of using secret keys to
achieve agreement, we use hash function and public keys to achieve
agreement. The argument here is that the content encrypted by public
keys will not be compromised easily, and we can use hash function to
commit values to be used as short term keys. Another particular
point of our work is to use a special tag including role names to
prevent role-mixup attack. Furthermore, we modify the $n$-party NSL
protocol in order to achieve  the higher agreement property
$nisynch$, which illustrates the practicability of the approach.

\subsection{Resilience of Secrecy Attack}

In \cite{02}, a tagging function for the transformation is provided.
We  recall that the function $\tau_c$ and the restricted one
$\tau_{c|S}$ defined as follows .

\par \noindent \textbf{Definition 12 (Tagging function)} Let $c\in Const$, $\tau_c: Term\rightarrow Term$, then for all $t,t_1,...,t_n\in Term$ :
\begin{equation}
\notag \tau_c(t)= \left\{
\begin{aligned}
&t,\qquad\qquad\qquad\text{if}\,t\,\text{atomic\,or\,long-term\,key},\\
&(\tau_c(t_1),\tau_c(t_2)),\qquad\quad\;\;\;\text{if}\,t=(t_1,t_2),\\
&\{\tau_c(t_1),c\}_{\tau_c(t_2)},\qquad\quad\;\;\text{if}\,t=\{t_1\}_{t_2},\\
&f(\tau_c(t_1),...,\tau_c(t_n)),c),\quad\text{if}\,t=f(t_1,...,t_n).
\end{aligned}
\right.
\end{equation}

$\tau_{c|S}$ denotes the modification of $\tau_{c}$ which restricts
the domain of $\tau_c$ to some set S of terms to avoid tagging
unnecessary terms.

 The transformation in Fig 5 shows how to ensure
AKCS of secrecy. Three messages are added: the first one is a
constant asking for a nonce, the second one contains an encrypted
nonce, and the third one contains the secrecy encrypted by the nonce
and the public key together. The last two works like encrypting
secrecy with two pair of keys, which the adversary at most
compromise either pair of them, thus achieving AKCS of secrecy for
both sides. Here we add different constant tags on message to ensure
the secrecy.

\begin{figure}[ht]
  \captionwidth{1.0\textwidth}
  \changecaptionwidth
  \centering
  \includegraphics[width=0.5\textwidth,height=0.4\textwidth,]{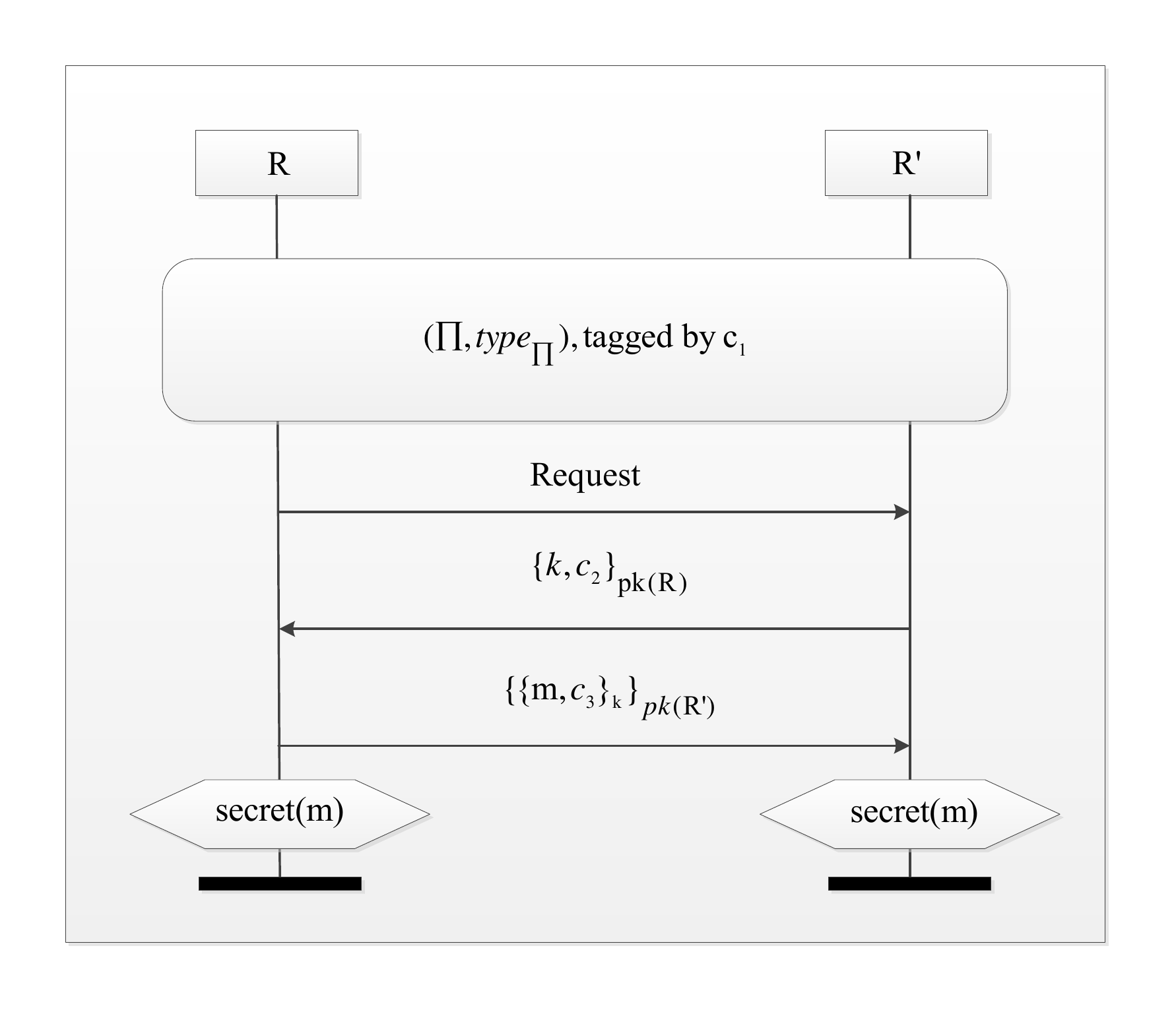}
  \\ {Fig.5. Transforming $\Pi$ for secrecy of m in both $R$ and $R'$.}
\end{figure}

Let $type_{TS(\Pi)}=type_\Pi$, $M=\{k,c_2\}_{pk(R)}$,
$N=\{\{m,c_3\}_k\}_{pk(R')}$, and

\begin{align}
\notag
&S_1=\langle send_{l_1}(R,R',Request),recv_{l_2}(R',R,M),\\
\notag &\quad\quad\quad send_{l_3}(R,R',N),claim_{l_4}(R,secret,m)\rangle\\
\notag
&S_2=\langle recv_{l_1}(R,R',Request),send_{l_2}(R',R,M),\\
\notag
&\quad\quad\quad recv_{l_3}(R,R',N),claim_{l_5}(R',secret,m)\rangle\\
\notag
&S=\{\{t\}_{t'}:type_\Pi(\{t\}_{t'})\cap type_{TS(\Pi)}(M)\neq \emptyset \}\\
\notag &\quad\cup \{\{t\}_{t'}:type_\Pi(\{t\}_{t'})\cap
type_{TS(\Pi)}(N)\neq\emptyset\}
\end{align}

The formal definition of the transformation is as follows.
\begin{equation}
\notag TS(\Pi)(x)= \left\{
\begin{aligned}
&\tau_{c_1|S}(\Pi(R))\cdot S_1,&if\;x=R,\\
&\tau_{c_1|S}(\Pi(R))\cdot S_2,&if\;x=R',\\
&\tau_{c_1|S}(\Pi(x)),&otherwise.
\end{aligned}
\right.
\end{equation}

Since no asymmetric long-term secret keys appear in accessible
position in a sent-message (a requirement stated at the end of
Section 2), and it can be proved \cite{02} that the adversary can
not reveal or infer the peers' asymmetric long-term secret key,
except the one the adversary knows through the given adversary rule.
The proof of the following proposition uses the fact that adversary
cannot forge the last message, therefore $m$ only appears in
accessible position of $\{m,c_3\}_k$. The secrecy of $m$ depends on
secrecy of $k$ and $pk(R')$, which cannot be compromised at the same
time. The reader is referred to \reportappendix for details.

\par \noindent \textbf{Proposition 1 (Secrecy by asymmetric encryption):}
\par \noindent Let $R,R'\in dom(\Pi)$ where $R\neq R'$. Let $A,A'$ an adversary which can compromise $R$ and $R'$ long-term secret key respectively. $c_1,c_2,c_3,Request\in Const$, $l_1,l_2,l_3,l_4,l_5\in Label$ and all of them are unequal and unused in $\Pi$. Let $k$, $n\in Fresh$, $m\in RoleTerm$ such that $n\sqsubseteq_{acc}m$ and $k$, $n$ all be unused in $\Pi$. If $(TS(\Pi),type_{TS(\Pi)})$ is a protocol and $type_{TS(\Pi)}=type_\Pi$ :
\begin{align}
\notag
(TS(\Pi),type_{TS(\Pi)})&\models_A claim_{l_4}(R,secret,m)\\
\notag (TS(\Pi),type_{TS(\Pi)})&\models_{A'}
claim_{l_5}(R',secret,m)
\end{align}
Then we can obviously get that $(TS(\Pi),type_{TS(\Pi)})\Rightarrow \neg SecrecyAttack(m,TS(\Pi),type_{TS(\Pi)})$.

\paragraph{\it Remarks}
The idea of adding messages to ensure secrecy is similar to that of
\cite{02}. The difference is that the purpose here is to  ensure
bilateral secrecy (i.e., no matter which key is compromised, the
secrecy of $m$ is guaranteed).

\subsection{On Substitution and Parallel Attack}
One way to prevent parallel attack is to  tag each message with a
hash function which includes all the previous variables. If the
adversary wants to disorganize one message between different
threads, it has to learn all the previous variables from both sides
which is very hard. In order to prevent substitution attack, we can
also take advantage of hash function by including new fresh and old
variables together in one hash function. Then the adversary cannot
forge a message using its own fresh because of the use of hash
functions. We use this technique in the following transformation
function and prove that the $commit$ property can be achieved with
AKC attacks.

The transformation in Fig 6 shows how to ensure AKCS of agreement.
We assume $m\in Fresh$ occurs in $\Pi$ and keeps secret. We add two
messages: the first one contains hash function of $m$ and $n$, where
$n$ is not used in the previous events. The second one is a response
using hash of $n$. The hash function here works like a signature,
where it takes use of $m$ or $n$'s secrecy to ensure that the
adversary can not forge the message.
\begin{figure}[ht]
  \captionwidth{1.0\textwidth}
  \changecaptionwidth
  \centering
  \includegraphics[width=0.5\textwidth,height=0.4\textwidth,]{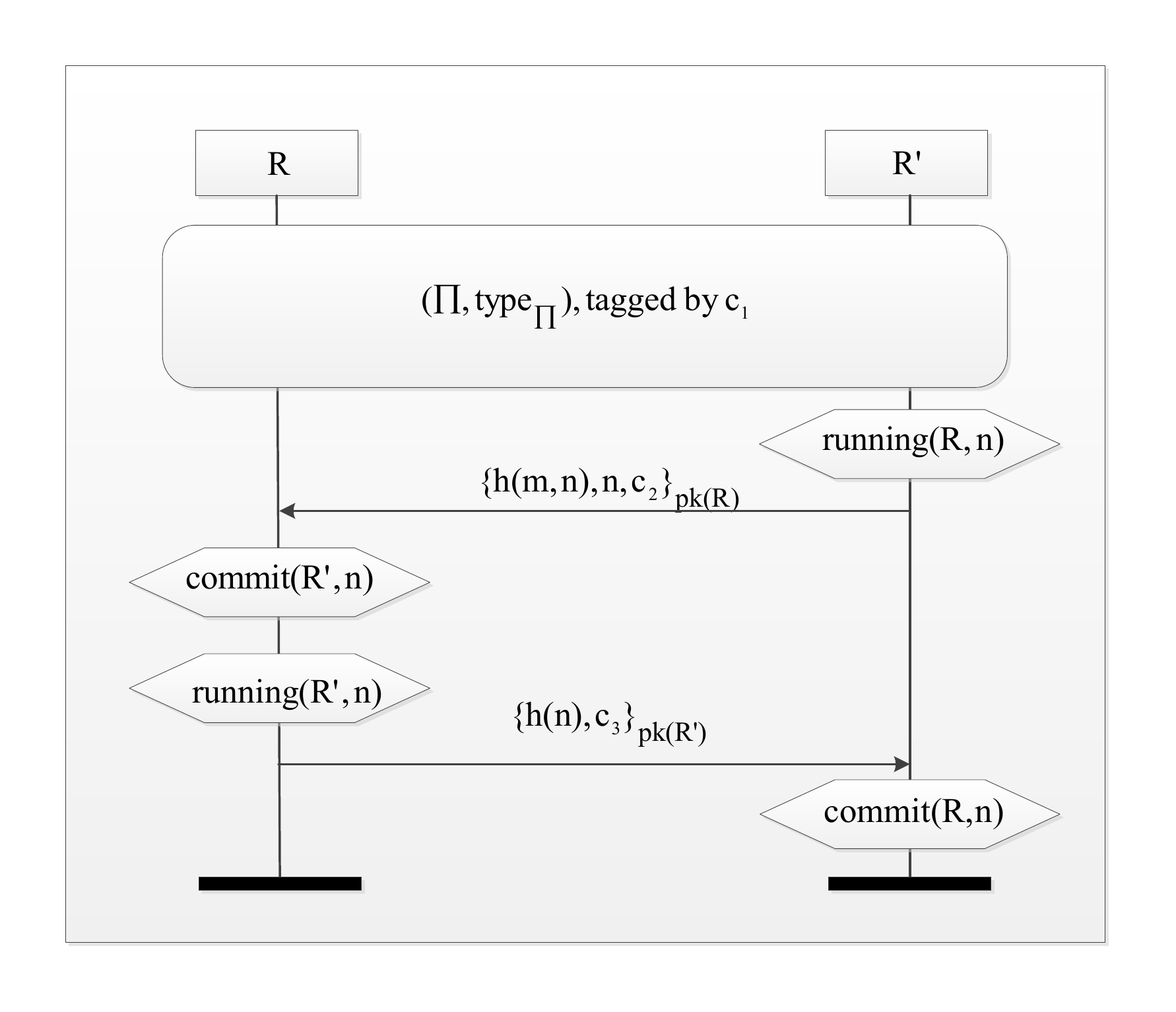}
  \\ {Fig.6. Transforming $\Pi$ for agreement on $n$ for both $R$ and $R'$.}
\end{figure}

Let $type_{TA(\Pi)}=type_\Pi$, $N=\{h(n), c_3\}_{pk(R')}$,
$M=\{h(m,n),n,c_2\}_{pk(R)}$, and

\begin{align}
\notag &S_1=\langle
recv_{l_2}(R',R,M), claim_{l_1}(R,commit,R',n),\\
\notag
&\qquad claim_{l_3}(R,running,R',n), send_{l_4}(R,R',N)\rangle\\
\notag &S_2=\langle claim_{l_1}(R',running,R,n),
send_{l_2}(R',R,M),\\
\notag
&\qquad recv_{l_4}(R,R',N), claim_{l_3}(R',commit,R,n)\rangle\\
\notag
&S=\{\{t\}_{t'}:type_\Pi(\{t\}_{t'})\cap type_{TA(\Pi)}(M)\neq \emptyset \}\\
\notag &\quad \cup \{\{t\}_{t'}:type_\Pi(\{t\}_{t'})\cap
type_{TA(\Pi)}(N)\neq\emptyset\}
\end{align}

The formal definition of the transformation is then as follows.
\begin{equation}
\notag TA(\Pi)(x)= \left\{
\begin{aligned}
&\tau_{c_1|S}(\Pi(R))\cdot S_1, &if\;x=R,\\
&\tau_{c_1|S}(\Pi(R))\cdot S_2,&if\;x=R',\\
&\tau_{c_1|S}(\Pi(x)),&otherwise.
\end{aligned}
\right.
\end{equation}

\par \noindent \textbf{Proposition 2 (Agreement by hashing):}
\par \noindent Let $R,R'\in dom(\Pi)$ such that $R\neq R'$.
Let $A,A'$ be adversaries which can compromise $R$ and $R'$
long-term secret key respectively. Let $l_1,l_2,l_3,l_4\in Label$
and $c_1,c_2,c_3\in Const$ all be different and unused in $\Pi$,
$m$, $n\in RoleTerm$, and $A$ an adversary such that $\forall s\in
RS(\Pi, type_{\Pi}, R, A)$, $AK_s \nvdash \sigma_{s,Test}(m)$. If
$(TA(\Pi),type_{TA(\Pi)})$ is a protocol and
$type_{TA(\Pi)}=type_\Pi$, then
\begin{align}
\notag
(TA(\Pi),type_{TA(\Pi)})&\models_A claim_{l_1}(R,commit,R',n)\\
\notag (TA(\Pi),type_{TA(\Pi)})&\models_{A'}
claim_{l_3}(R',commit,R,n)
\end{align}


The reader is referred to \reportappendix for a proof. This kind of
transformation is resilient against substitution and parallel
attack.

\par \noindent \textbf{Corollary 1 (Resilience of Substitution Attack)}
If the original protocol is resilient against substitution attack,
then the modified protocol keeps this property:
\begin{align}
\notag &\neg SubAttack(\Pi, type_{\Pi}) \Rightarrow\neg
SubAttack(TA(\Pi), type_{TA(\Pi)})
\end{align}

\par \noindent \textbf{Corollary 2 (Resilience of Parallel Attack)} If
the original protocol is resilient against parallel attack, then the
modified protocol keeps this property:
\begin{align}
\notag &
\neg ParallelAttack(\Pi, type_{\Pi})\\
\notag &\qquad\Rightarrow \neg ParallelAttack(TA(\Pi),
type_{TA(\Pi)})
\end{align}

The two corollaries is used to transform protocol inductively. We can assume a protocol to be empty at first, then add each message based on proposition 2 to ensure agreement. The reader is referred to \reportappendix for the  proofs of the
corollaries.

\subsection{Resilience of Role Mixup Attack}

For preventing role-mixup attacks, we find a special kind of tags,
which contain all role names encrypted by secret keys, very useful.
Let $t,t_1,...,t_n\in Term$ be terms. Let $AR(x)=\{dom(\Pi)\setminus
x\}_{sk(x)}$, the tagging function $\upsilon_{x}(t)$ is defined as
follows.

\begin{equation}
\notag \upsilon_{x}(t)= \left\{
\begin{aligned}
&t,\qquad\qquad \text{if}\;t\;\text{atomic\;or\;a\;long-term\;key,}\\
&(\upsilon_x(t_1),\upsilon_x(t_2), AR(x)),\quad\text{if}\;t=(t_1,t_2),\\
&\{\upsilon_x(t_1),AR(x)\}_{t_2},\qquad\quad\text{if}\;t=\{t_1\}_{t_2}.
\end{aligned}
\right.
\end{equation}

Let $v_x: Term \rightarrow Term$ extends to $Event^* \rightarrow
Event^*$ by replacing all terms in the event sequence accordingly.
This will then provide a transformation function $TR(\Pi)$ such that
$TR(\Pi)(x) = \upsilon_{x}(\Pi(x))$.


Assume that the content of every message is composite (in contrast
to atomic terms) and any $send$-event has response. Then this
transformation is helpful for preventing role-mixup attack.
The reason is that, if we consider agent names as fresh values, then
based on proposition 11 in \cite{02}, every two parties which
communicated with each other agree on all the agent names. Because
the communication among parties can form a strongly connected graph,
so all parties agree on the agent names. Then if there is role-mixup
attack, there exists reachable state $s$ such that either $Replace$
or $Termin$ function holds. Since each party has agreed on which
agent instantiated which role, replacement or forgery can detected
by the agents.

In the following, we apply this technique together with the
transformations provided in Propositions 1 and 2 to achieve
$nisynch$-property of multi-party protocols.


\par \noindent \textbf{AKCS in Multi-Party Authentication Protocols}
Multi-party protocols are more vulnerable to AKC attacks as a result
of complicated communications among parties. We consider a family of
multi-party NSL protocols, which are brought up by  \cite{14}. The
protocols  are vulnerable to AKC attacks.
%
Let the protocols be denoted $(\Pi_p,type_{\Pi_p})$ where $p$
denotes the number of parties in the particular protocol.

The approach for the transformation is as follows. We first modify
messages between each pair of parties, and add hash function tags in
them to prevent substitution and parallel attack. Then we combine
the messages between each pair to form a new protocol, and finally
add $AR(x)$ tags to prevent role-mixup attack.
Let $n_0,...,n_{p-1}\in Fresh, R_0,...,R_{p-1}\in \Role$, and $$
\begin{array}{l}
\notag M_A(i)=\{\{n_0,...,n_i,AR(R_i)\}_{sk(R_i)}\}_{pk(R_{i+1})}
\\ \notag
M_B(i)=\{h(n_0,...,n_i,R_0,...,R_{p-1}),n_1,...,n_i\}_{pk(R_0)} \\
\notag
M_C(i)=\{h(n_{i+1},...,n_{p-1}),n_{i+2},...,n_{p-1}\}_{pk(R_{i+1})})
\end{array}
$$

Then we define the $i$'th protocol message, for $0\leqslant i
<2p-1$, by
\begin{equation}
\notag Msg(i)= \left\{
\begin{aligned}
&M_A(i),\quad if\;0\leqslant i<p-1,\\
&M_B(i),\quad if\;i=p-1,\\
&M_C(i),\quad if\;p-1<i<2p-1.
\end{aligned}
\right.
\end{equation}

Here we simplify the tag function $\nu_x(t)$, because it is
sufficient to tag only the first round of communication in one
accessible position. Furthermore, we encrypt fresh with secret key
in $M_A$ to ensure the agreement. Let
$l_0,...,l_{2p-1},m_0,...,m_{p-1}$ be labels, and
$S_1$ and $S_2$ be defined as follows.
\begin{align}
\notag
S_1=&\langle send_{l_0}(R_0,R_1,Msg(0)),\\
\notag
&recv_{l_{p-1}}(R_{p-1},R_0,Msg(p-1)),\\
\notag
&send_{l_p}(R_0,R_1,Msg(p-1)),\\
\notag
&claim_{m_0}(R_0,nisynch)\rangle\\
\notag
S_2(i)=&\langle recv_{l_{i-1}}(R_{i-1},R_{i}Msg(i-1)),\\
\notag
&send_{l_i}(R_i,R_{i+1},Msg(i)),\\
\notag
&recv_{l_{i+p}}(R_{i-1},R_i,Msg(i+p)),\\
\notag &claim_{m_i}(R_i,nisynch)\rangle
\end{align}

The modification of a such a protocol $(\Pi_p,type_{\Pi_p})$ is as
follows (with $type_{\Pi_p}$ keeps unchanged).
\begin{equation}
\notag TM(\Pi_p)(x)= \left\{
\begin{aligned}
&S_1, &if\;x&=R_0,\\
&S_2(i), &if\;x&=R_i\;(0<i\leqslant p-1).
\end{aligned}
\right.
\end{equation}

This transformed protocol has the same structure as the original one
with each message replaced by the given ones. The correctness with
respect to the $nisynch$ claim is stated in the following
proposition and proved  by using the fact that, the message
encrypted by asymmetric secret key or contain hash functions on
secret nonce can achieve agreement between two parties. The reader
is referred to \reportappendix for a proof.

\par \noindent \textbf{Proposition 3 (Multi-party NSL agreement):}
\par \noindent
Let$(TM(\Pi_p),type_{TM(\Pi_p)})$ be the transformed protocol, with
$dom(TM(\Pi_p))=\{R_0,...,R_{p-1}\}$. Let $A_0,...,A_{p-1}$ be
adversaries which can compromise the respective long-term secret key
of $R_i$. Let $\gamma(x)=claim_{m_x}(R_x,nisynch)$. Then
\begin{align}
\notag TM(\Pi_p),type_{TM(\Pi_p)})&\models_{A_i} \gamma(i) \; for \;
i=0,...,p-1.
\end{align}

\section{Case Studies}

Many protocols are vulnerable under AKC attacks, with examples shown
in Section 3. We have applied the above techniques to enhance the
security level of such protocols.
In accordance with the transformation provided in Propositions 1, 2,
we transform these protocols into AKCS ones. Table 1 shows part of
the results of experiments using the Scyther tool \cite{Scyther}
after that we have applied the transformation scheme.
'--' means the property is not required for the protocol. For
$\surd$ we means the property holds for each party in the protocol
(after the transformation).

\begin{table}[htbp]
\centering
 \caption{\label{tab:summary of protocol}Protocol Experiment}
 \begin{tabular}{lcl}
  \toprule
  {protocol} & {secrecy} & {nisynch} \\
  \midrule
 Bilateral Key Exchange & kir$(\surd)$ & $\surd$ \\
 CCIT-ban1  & Ya$(\surd)$ & $\surd$ \\
 CCIT-ban3  & Ya,Yb$(\surd)$ & $\surd$ \\
 isoiec-9798-3-5 & -- & $\surd$ \\
 NSL & ni,nr$(\surd)$ & $\surd$ \\
 PKMV2RSA & prepak$(\surd)$ & $\surd$ \\
 Kerberos & Kr$(\surd)$ & $\surd$\\
 TMN & ST$(\surd)$ & $\surd$\\
 Splice/AS & N2$(\surd)$ & $\surd$\\
 Cardholder-Registration & PAN$(\surd)$ & $\surd$\\
  \bottomrule
 \end{tabular}
\end{table}

In the following, we demonstrate how the three practical protocols,
PKMV2RSA, Kerberos and Cardholder-Registration protocols, are
transformed. We give the original model of these protocols, point
out the AKC attack on authorization and secrecy in them and
transform the protocol based on the propositions.

\subsection{PKMV2RSA}
PKMV2RSA \cite{15} is a subprotocol of WiMAX, which known as a
wireless access system to deliver the "last mile" wireless broadband
access. The subprotocols are used for authentication, key
management, and secure communication. Among them, PKMV2RSA
authenticates the base station (BS) and mobile station (MS) and
establishes a shared secret which is used to secure the exchange of
traffic encryption keys (TEKs). There are six messages in all, but
since the secrecy of TEKs depends on the secrecy of prepak, and the
last three messages is resilient against AKC attack, then we only
need to look at the first three messages. The protocol proceeds as
follows:
\begin{align}
& \notag 1.\;MS \rightarrow BS : \{msrand,said,MS\}_{sk(MS)} \\
& \notag 2.\;BS \rightarrow MS : \{msrand,bsrand,\\
& \notag
\qquad\qquad\qquad\;\;\{prepak,MS\}_{pk(MS)},BS\}_{sk(BS)}\\
& \notag 3.\;MS \rightarrow BS : \{bsrand\}_{sk(MS)}
\end{align}

The secrecy of prepak is based on the secrecy of mobile station's
long-term secret key $sk(MS)$.  Then there is AKC attack on secrecy
of TEKs and agreement of both sides. We implement the protocol by
using $said$ to encrypt $prepak$ in message 2, and add hash function
on message 3, which is an example of the transformation scheme of
Propositions 1 and 2. The modified protocol is as follows.
\begin{align}
& \notag 1.\;MS \rightarrow BS : \{msrand,\{said\}_{pk(B)},MS\}_{sk(MS)} \\
& \notag 2.\;BS \rightarrow MS : \{msrand,bsrand,\\
& \notag
\qquad\qquad\quad\quad\;\;\{\{prepak\}_{said},MS\}_{pk(MS)},BS\}_{sk(BS)}\\
& \notag 3.\;MS \rightarrow BS :
\{h(bsrand,msrand,prepak)\}_{sk(MS)}
\end{align}

As shown in Table 1, this modified protocol satisfies the
$nisynch$-property, the claim on the secrecy of $prepak$ holds.

\subsection{Kerberos}
Kerberos \cite{16} is designed to authenticate clients to multiple
networked services. PKINIT, an extension of Kerberos 5, is modified
to allow public-key authentication. The basic Kerberos has four
parties: Client (C), whose goal is to authenticate itself to various
application servers; Kerberos Authentication Server (KS), who
provide "ticket-granting ticket" (TGT); Ticket-Granting Server (TS),
who is presented TGT and then provide "server ticket" (ST) to
client. ST is the credential that client uses to authenticate
herself to the application server. Since role C talk to KS, TS and S
separately, we can divide the protocol to three two-party parts. We
show the first part below:
\begin{align}
& \notag 1. \; C \rightarrow KS : \{Tc,n,C,KS,TS\}_{sk(C)} \\
& \notag 2. \; KS \rightarrow C : \{\{k,H(C,TS,\{Tc,n,C,KS,TS\}_{sk(C)}),\\
&\notag \quad TGT\}_{sk(KS)}\}_{pk(C)}, \{AK,Tk,TS\}_k
\end{align}

The main issue is to ensure secrecy of ST before client sends it to
the server,  and the secrecy of ST depends on secrecy of AK, which
depends on secrecy of k. However, k can be revealed if the intruder
knows sk(C) and it is easy for the intruder to fake a message 2 and
sent it to KS. Therefore we use Propositions 1 and 2 to modify
message 2 as follows.
\begin{align}
\notag
&\{\{\{k\}_n,H(C,TS,n,\{Tc,n,C,KS,TS\}_{sk(C)}),\\
\notag &C,TGT\}_{sk(KS)}\}_{pk(C)},\{AK,n,Tk,TS\}_k
\end{align}

Then part 1 can achieve both secret and nisynch property. The other
two parts can be modified similarly.

\subsection{Cardholder-Registration}

Cardholder-Registration protocol \cite{17} is the first part of SET
protocol in online purchase. It comprises three message exchange
between the cardholder and a certificate authority. In the first
exchange, the cardholder requests registration and is given the
certificate authority's public keys. In the second exchange, the
cardholder supplies his credit card number (PAN) and receives an
application form for the bank that issued his credit card. In the
third exchange, the cardholder returns the completed application
form and delivers his public signature key and supplies a
CardSecret. This process is as follows.
\begin{align}
& \notag 1. \; C \rightarrow CA : \{C,Nc1\}_{pk(CA)}\\
& \notag 2. \; CA \rightarrow C : \{C,H(Nc1)\}_{pk(C)} \\
& \notag 3. \; C \rightarrow CA : \{C,Nc2,H(PAN)\}_{c1},\{c1,PAN\}_{pk(CA)}\\
& \notag 4. \; CA \rightarrow C : \{C,Nc2,Nca\}_{pk(C)}\\
& \notag 5. \; C \rightarrow CA : \{C,Nc3,c2,pk(C),\{H(C,Nc3,c2,pk(C),\\
&\notag\quad
PAN, NsecC)\}_{sk(C)}\}_{c3},\{c3,PAN,NsecC\}_{pk(CA)}\\
& \notag 6. \; CA \rightarrow C : \{C,c3,CA,NsecCA\}_{c2}
\end{align}

The protocol is not secure: the secrecy PAN, NsecC, and NsecCA will
be revealed if C or CA's long-term secret key is compromised. It
also fails to reach agreement: message 3, 4 or 5, 6 contains no
previously received messages, and is thus vulnerable to parallel
attacks.
We can modify the protocol by inserting a new nonce Nc4 to encrypt
PAN and NsecC  and adding hash tags in each message to guarantee
nisynch property. The modified protocol is as follows.
\begin{align}
& \notag 1. \; C \rightarrow CA : \{C,Nc1\}_{pk(CA)}\\
& \notag 2. \; CA \rightarrow C : \{C,H(Nc1,Nc4),Nc4\}_{pk(C)} \\
& \notag 3. \; C \rightarrow CA : \{C,Nc2,H(PAN)\}_{c1},\{\{c1,PAN\}_{Nc4},\\
&\notag
\quad H(C,Nc2,Nc1,c1)\}_{pk(CA)}\\
& \notag 4. \; CA \rightarrow C : \{C,Nc2,\{Nca\}_{Nc1},\\
&\notag
\quad H(Nc2,Nca,Nc1)\}_{pk(C)}\\
& \notag 5. \; C \rightarrow CA : \{C,Nc3,c2,pk(C),\{H(C,Nc3,c2,\\
&\notag
\quad \; pk(C),PAN,Nc2,Nca,Nc1,NsecC)\}_{sk(C)}\}_{c3},\\
&\notag
\quad\{c3,PAN,\{NsecC\}_{Nc4}\}_{pk(CA)}\\
& \notag 6. \; CA \rightarrow C : \{\{C,c3,CA,\{NsecCA\}_{Nc1},\\
&\notag \quad H(Nc2,Nca,Nc1,NsecC,NsecCA)\}_{sk(CA)}\}_{c2}
\end{align}

The modification guarantees the secrecy of $PAN$ and the
$nisynch$-property.

\section{Concluding Remarks}

This paper gives an analysis of AKC attacks and provides solutions
to enhance the level of security. We consider four types of AKC
attacks and give the definition of these types. Then based on the
attack types, we provide techniques for transformation of protocols.
A guiding principle in designing security protocol under potential
AKC attacks is using short-term keys to ensure secrecy, hash
functions to maintain agreement and role names to prevent role-mixup
attack. We have applied the techniques to the transformation of
practical protocols and have used the verification tool $Scyther$ to
show that the modified protocols have achieved higher level of
security.


\section{Appendix}

\subsection{Proofs}

Before presenting the proofs of the propositions and corollaries, we
present 3 lemmas.
Lemma 1 states that if some term $t$ is secret before some event and
no parts of $t$ occur in accessible positions in the later events,
then it keeps secret at the end of the sequence of the events. Lemma
2 states that a term encrypted by a secret nonce must have been sent
by an agent, because no derivation of the term from $AK_s$ is
possible. Lemma 3 states a similar property with a hashed term.

\bigskip

\par \noindent \textbf{Lemma 1}:
Let $s,s'$ be states such that $s' \rightarrow^* s$. Suppose that
$last(tr_{s'})=(tid,e')$ and $last(tr_s)=(Test,e)$, where $last()$
denotes the last element of a sequence. Suppose that
$t\sqsubseteq_{acc}cont(e')$. If for all $e''$ such that $label(e')
< label(e'') < label(e)$, each $t''\sqsubseteq_{acc}cont(e'')$ has
never been used before $s'$, then
\begin{align}
\notag AK_{s'}\nvdash \sigma_{s',Test}(t) \Rightarrow AK_{s}\nvdash
\sigma_{s,Test}(t).
\end{align}

\par \noindent \textbf{Proof of Lemma 1:}
Using the execution rules and adversary rules, we have $AK_s =
AK_{s'} \cup K$ where $K$ denotes newly added adversary knowledge
between $l'$ and $l$. We want to prove that $AK_{s}\nvdash
\sigma_{s,Test}(t)$. We have $AK_{s'} \nvdash \sigma_{s,Test}(t)$,
and $t\not\sqsubseteq_{acc}cont(e'')$ for every $e''$ that appears
between $s'$ and $s$, then we get $t\not\sqsubseteq_{acc}K$. Because
for each term $t''$ that we get from accessible position of $e''$,
$t''$ has never been used before, thus $K$ is not helpful in
deducing $t$. Then we get $AK_{s}\nvdash \sigma_{s,Test}(t)$.

\bigskip

\par \noindent \textbf{Lemma 2}:
Suppose that  $n=\{m^{\#tid}, c\}_k$ with $c\in Const$, $k\in
Fresh$, $m,n \in Runterm$, $tid \in TID$. Let $s$ be a reachable
state such that $AK_s\nvdash \sigma_{s,Test}(k^{\#Test})$. If
$tr_s\cdot\langle(Test,recv_l(a,b,\{m\}_{pk(b)}))\rangle\in
Traces(\Pi, type_{\Pi})$ for some $a,b,l$, then
\begin{align}
\notag \exists (tid,e')\in tr_s.(evtype(e')=send \land
n\sqsubseteq_{acc}cont(e')).
\end{align}

\par \noindent \textbf{Proof of Lemma 2:}
Since $AK_s\nvdash k$, no derivation of $\sigma_{s,Test}(\{m,c\}_k)$
can end in a composition step, which implies that
$m\sqsubseteq_{acc} AK_{s}$ by Lemma 6 of \cite{02}. Therefore there
exists $tid'\in TID$,  $e\in RunEvent$ such that $(tid',e)\in tr_s$,
$evtype(e)=send$, and $n\sqsubseteq_{acc}cont(e)$.

\bigskip

\par \noindent \textbf{Lemma 3}:
Suppose that $m = ( h(n,t),t^{\#tid})$ with $h\in Func$, $n,t\in
Fresh$, $tid \in TID$. Let $s$ be a reachable state such that
$AK_s\nvdash \sigma_{s,Test}(n^{\#Test})$. If
$tr_s\cdot\langle(Test,recv_l(a,b,\{m,c\}_{pk(b)}))\rangle\in
Traces(\Pi, type_{\Pi})$ for some $c\in Const$ and $a,b,l$, then
\begin{align}
\notag \exists (tid,e)\in tr_s.(evtype(e)=send \land
m\sqsubseteq_{acc}cont(e)).
\end{align}

\par \noindent \textbf{Proof of Lemma 3:} Since $AK_s\nvdash n^{\#Test}$, and $t^{\#tid}$
has  first appear in $m$, we get $AK_s\nvdash m$. If $m$ can be
forged by adversary, then it has to know $n^{\#Test}$ which is not
accessible by adversary. That means no derivation of
$\sigma_{s,Test}(m)$ can end in a composition step. Then we get
$m\sqsubseteq_{acc}AK_s$ by Lemma 6 in \cite{02}. Therefore there
exists $tid'\in TID$,  $e\in RunEvent$ such that $(tid',e)\in tr_s$,
$evtype(e)=send$, and $m\sqsubseteq_{acc}cont(e)$.

\bigskip

\par \noindent \textbf{Proof of Proposition 1:}

Let $\Pi'=TS(\Pi)$.

(1) We prove $(\Pi',type_{\Pi'})\models_A claim_{l_4}(R,secret,m).$

Let $s \in RS(\Pi',type_{\Pi'},R,A)$ such that

$(Test,\sigma_{s,Test}(claim_{l_4}(R,secret,m^{\#Test})))\in tr_s$.

The goal is to prove that $AK_s\nvdash \sigma_{s,Test}(m^{\#Test})$.

Let $N = \{m,c_3\}_k\in RoleTerm$.

According to Proposition 10 of \cite{02}, we get $AK_s\nvdash N$.

Since $m\sqsubseteq_{acc}N$ appears first time in $N$, we have
$AK_s\nvdash \sigma_{s,Test}(m^{\#Test})$.

(2) We prove $(\Pi',type_{\Pi'}) \models_{A'}
claim_{l_5}(R',secret,m)$.

Let $s\in RS(\Pi',type_{\Pi'},R',A')$ such that

$(Test,\sigma_{s,Test}(claim_{l_5}(R',secret,m)))\in tr_s$.

The goal is to prove that $AK_s\nvdash
\sigma_{s,Test}({m}^{\#tid})$.

At step 1, we want to prove $AK_s\nvdash
\sigma_{s,Test}({k}^{\#Test})$.

Let $s'\in RS(\Pi',type_{\Pi'},R',A')$ such that $s'\rightarrow^*s$.

Let $tid'\in TID$ and $e'\in Event$ such that
$(tid',e')=last(tr_{s'})$, $evtype(e')=send$,
$k^{\#Test}\sqsubseteq_{acc}cont(e')$.

According to Proposition 10 of \cite{02}, we have $AK_{s'}\nvdash
\sigma_{s',Test}(k^{\#Test})$.

By Lemma 1, we get $AK_s\nvdash \sigma_{s,Test}({k}^{\#Test})$.

 By Lemma 2, there exists $tid'$,  $e$ such that
$(tid',e)\in tr_s$, $evtype(e)=send$, $m\sqsubseteq_{acc}cont(e)$.

Assume that $l'\neq l_5$, then $e$ is an instance of a tagged step
of $\Pi$, such that there exist $t'\in RoleTerm$ and
$\sigma_{s,tid'}(t'^{\#tid'})=cont(e)$ and
$send_{l'}(\cdot,\cdot,t')\in \tau_{c_1|S}(\Pi(role_{s}(tid)))$.

Then there exists $\{t_0\}_{t_1}\in S$ such that

$\sigma_{s,tid'}(\tau_{c_1}((\{t_0\}_{t_1})^{\#tid}))=\sigma_{s,Test}(\{m^{\#tid},c_3\}_k)$.

This implies that $c_1= c_3$ and contradicts the conditions of the
transformation.

 Hence $l'=l_5$.

 Since $AK_s\nvdash
\sigma_{s,Test}({k}^{\#Test})$ and $m^{\#tid}$ appears in $e$ first
time, according to Proposition 10 of \cite{02}, we have that
$m^{\#tid}$ is only accessible in the set $AK_s$ as a subterm of the
term $\sigma_{s,Test}(\{m,c_3\}_k)$.

Since we have proved that $AK_s\nvdash
\sigma_{s,Test}({k}^{\#Test})$, we have $AK_s\not\in
\sigma_{s,Test}({m}^{\#Test})$.

\bigskip

\par \noindent \textbf{Proof of Proposition 2:}

Let $\Pi'=TS(\Pi)$.

(1) We prove $(\Pi',type_{\Pi'})\models_A
claim_{l_1}(R,commit,R',n)$.

Let $s\in RS(\Pi',type_{\Pi'},R,A)$ such that

$(Test,\sigma_{s,Test}(claim_{l_1}(R,commit,R',n^{\#tid})))\in
tr_s$.

We prove that the corresponding running claim holds.

Let $t=\sigma_{s,Test}(h(m,n),n^{\#tid})$.

Since $AK_s\nvdash\sigma_{s,Test}(n^{\#tid})$, by Lemma 3,  there
exists $tid'$, $e$ such that $(tid',e)\in tr_s$, $evtype(e)=send$,
 $t\sqsubseteq_{acc}cont(e)$.

Assume that $l'\neq l_1$, then $e$ is an instance of a tagged event
of $\Pi'$.

Then there is a $\{t_0\}_{t_1}\in S$ and
$\sigma_{s,tid'}(\tau_{c_1}((\{t_0\}_{t_1})^{\#tid}))=\sigma_{s,Test}(\{h(m,n),n^{\#tid},c_2\}_{pk(R)})$,
which contradicts $c_1\neq c_2$.

Hence $l'=l_1$. Therefore the running claim holds.

(2) We prove $(\Pi',type_{\Pi'})\models_{A'}
claim_{l_3}(R,commit,R',n)$.

Let $s\in RS(\Pi',type_{\Pi'},R,A)$ such that

$(Test,\sigma_{s,Test}(claim_{l_3}(R',commit,R,n^{\#Test})))\in
tr_s$.

According to Proposition 10 of  \cite{02} and Lemma 1, we get
$AK_s\nvdash \sigma_{s,Test}(n^{\#Test})$.

The rest of the proof is similar to the above one, in which we use
Lemma 3 to prove that the corresponding running claim holds.

\bigskip

\par \noindent \textbf{Proof of Corollary 1:} If either $R$ or $R'$
long-term secret key is compromised, from the proof of Proposition
2, we know that $\exists i,j,i',j'\in N$, $i<j$, $i'<j'$, $a,b\in
\Agent$, $tid\in TID$ and a reachable state $s$ such that

$tr_{s,i}=\sigma_{s,Test}(send_{l_2}(b,a,h(m,n),n^{\#tid})),$

$tr_{s,j}=\sigma_{s,tid}(recv_{l_2}(b,a,h(m,n),n^{\#tid})),$

$tr_{s,i'}=\sigma_{s,tid}(send_{l_4}(a,b,h(n))),$ and

$tr_{s,j'}=\sigma_{s,Test}(recv_{l_4}(a,b,h(n)))$.

Then according to the precondition,  we have that for each label
$l\in prec(l_3)$, $\exists i,j\in N$, $i < j$, $tid_1,tid_2\in TID$,
such that $Match(a,tid_1,b,tid_2)$ and
$ev(tr_{s,i})=send_{l}(a,b,m)\wedge ev(tr_{s,j})=recv_{l}(a,b,m)$.

This has violate the definition of substitution attack. Therefore
the conclusion is correct.

\bigskip

\par \noindent \textbf{Proof of Corollary 2:} Since we have proved there exists $tid_1,tid_2\in TID$ such that $Match(a,tid_1,b,tid_2)$ for corresponding send and recv events, which also violates the definition of parallel attack, then the conclusion is correct.

\bigskip

\par \noindent \textbf{Proof of Proposition 3:}

Let $p$ be arbitrary given, and let $\Pi'=TM(\Pi_p)$.

(1) First, we prove that, for a reachable state $s$, $AK_s\nvdash
\sigma_{s,Test}(n_k)$. Since $\sigma_{s,Test}(n_k)$ appears first time in the $send$-event
of $R_k$, and each accessible position where $\sigma_{s,Test}(n_k)$
appears is encrypted by $pk(R_s)$ where $(s\neq k)$, and $A_k\nvdash
sk(R_s)$, therefore $AK_s\nvdash \sigma_{s,Test}(n_k)$.

(2) Then we prove that, each agent has the same assumption  of agent
names with others. For adversary $A_0$, if any agent has different
assumption of agent names with $\sigma_{s,Test}(R_0)$, because
$AK_s\nvdash sk(\sigma_{s,Test}(R_x))(x\neq 0)$ and agent names were
transmitted between $R_1$ and $R_{p-1}$ by secret key, then
$\sigma_{s,Test}(R_{p-1})$ has different assumption with
$\sigma_{s,Test}(R_0)$. Since $AK_s\nvdash \sigma_{s,Test}(n_0)$,
then $\sigma_{s,Test}(Msg(p-1))$ cannot end in a compositional step,
then $\sigma_{s,Test}(R_0)$ will find that he has different
assumption with others, and terminates the protocol, which violates
the premise of $nisynch$ property. Therefore, for adversary $A_0$,
all agent has the same assumption of agent names. The proof for
other adversary $A_x$ is similar.

(3) We look at the role $R_k$  with $A=A_k$ for $0<k\leqslant p-1$.

Let $s\in RS(\Pi', type_{\Pi'}, R_k, A)$ with a position $q_i$ such
that:
$tr_{s,q_i}=(Test, claim_{m_k}({R_k}^{\#Test},nisynch)).$

Let $q_{j-1},q_{j},q_{j+k}$ be positions such that $0\leqslant
q_{j-1}<q_{j}<q_{j+k}< q_i$.

Let $a=\sigma_{s,Test}(R_{k-1})$, $b=\sigma_{s,Test}(R_{k})$, and
$c=\sigma_{s,Test}(R_{k+1})$. Then
\begin{align}
\notag
&ev(tr_{s,q_{j-1}})=recv_{l_{k-1}}(a,b,Msg(k-1))^{\#Test},\\
\notag
&ev(tr_{s,q_{j}})=send_{l_k}(b,c,Msg(k))^{\#Test},\\
\notag &ev(tr_{s,q_{j+k}})=recv_{l_{k+p}}(a,b,Msg(k+p))^{\#Test}.
\end{align}

We want to  prove that there are positions $q_{j'-1}$, $q_{j'}$,
$q_{j'+k}$ and  $tid_1,tid_2\in TID$, such that $q_{j'-1}<q_{j-1}$,
$q_{j}<q_{j'}$, $q_{j'+k}<q_{j+k}$, and
\begin{align}
&ev(tr_{s,q_{j'-1}})=send_{l_{k-1}}(a,b,Msg(k-1))^{\#tid_1},\\
&ev(tr_{s,q_{j'}})=recv_{l_k}(b,c,Msg(k))^{\#tid_2},\\ &ev(tr_{s,q_{j'+k}})=send_{l_{k+p}}(a,b,Msg(k+p))^{\#tid_1}.
\end{align}

(3a) First we look at label $l_{k+p}$. For adversary $A_k$, we have proved $AK_s\nvdash
\sigma_{s,Test}(n_k)$. We use Lemma 3 to establish position
$q_{j'+k}$ and $tid_1$ such that $q_{j'+k}< q_{j+k}$ and the
equalities
$ev(tr_{s,q_{j'+k}})=send_{l_{k+p}}(a,b,Msg(k+p))^{\#tid_1}$.

(3b) Then we look at label $l_{k-1}$. For adversary $A_k$,  since $AK\nvdash
sk(R_{i})(i\neq k)$, and $pk(R_k)$ can not be replaced as $AR(x)$
has determined the agent, then no derivation of $\sigma_{s,
tid_1}(send_{l_{k-1}}(a,b,MsgA(k-1)))$ from $AK_s$ can end in a
composition step. Then there exists $q_{j'-1}< q_{j-1}$ such that
$ev(tr_{s,q_{j'-1}})=send_{l_{k-1}}(a,b,Msg(k-1))^{\#tid_1}$.

(3c) At last we look at label $l_k$.  We have proved that $R_k$ has agree on
$n_k$ by receiving message $Msg(k+p)$. Then we deduce that $R_{k+1}$
has $Msg(k)$ which has $n_k$ in accessible positions. Since
$AK\nvdash sk(R_{i})(i\neq k)$, then there exists $q_{j'}$ and
$q_{j}<q_{j'}$ such that
$ev(tr_{s,q_{j'}})=recv_{l_k}(b,c,Msg(k))^{\#tid_2}$.

(4) At last, we look at the role $R_0$  with $A=A_0$.

Let $s\in RS(\Pi', type_{\Pi'}, R_0, A)$. Since we already have
$AK_s\nvdash \sigma_{s,Test}(n_0)$, then by Lemma 3, we have that
there exists a send-event corresponding to
$recv_{l_{p-1}}(R_{p-1},R_0,Msg(p-1))$. Since it is the only
recv-event for $R_0$, we are done with the proof.

\subsection{Algorithms}

In this subsection, we present algorithms for the transformation
based on the transformation scheme provided in Section 4.

\subsubsection{Protocol Syntax}


For practical reasons, we make restrictions on the protocol syntax.
We require that the content in a message has some fixed structure.
The terms in a protocol are organized such that role names appears
first, and then fresh names, then hash functions, etc. Each fresh
appears accessible only once in a message. The role in pk(r) should be the
responder, and the role in sk(r) should be the initiator. Terms in
the original message should not be encrypt by fresh names, but it
can be encrypt after the transformation. The protocols are defined
as follows.
\begin{align}
\notag protocol&::=mess^*,claim^*\\
\notag
mess&::=Role,Role,tm,tmp,tms,tmps,tmsp\\
\notag
tm&::=\varepsilon\;|\;tmr,tmf,tmh,tmn\\
\notag tmp&::=\varepsilon\;|\;\{tm\}_{pk(tmr)}\\
\notag tms&::=\varepsilon\;|\;\{tm\}_{sk(tmr)}\\
\notag tmh&::=\varepsilon\;|\;h(tmf)\\
\notag tmr&::=Role^*\\
\notag tmf&::=Fresh^*\\
\notag tmn&::=\varepsilon \;|\;\{tmf\}_{Fresh}\\ \notag tmps&::=\varepsilon\;|\;\{tm,tmp\}_{sk(tmr)}\\
\notag tmsp&::=\varepsilon\;|\;\{tm,tms\}_{pk(tmr)}\\
\notag claim&::=(Role,secret,Fresh)^*\\
\notag
&\quad |\;\;(Role,commit,Role,Fresh)^*\\
\notag &\quad |\;\;(Role,nisynch)^*
\end{align}


\subsubsection{Functions}

For events and messages, a set of operations are defined. $cn$
collects fresh names in messages, $chn$ collects fresh names
appearing in hash functions, $cs$ collects fresh names in
secrecy-claims, $cc$ collects fresh names in commit-claims.
\begin{align}
\notag
&cn(mess)=\\
\notag
&\{f\in Fresh\:|\: \exists s\in tmf.(s\sqsubseteq_{acc}mess\wedge f\sqsubseteq_{acc}s)\})\\
\notag
&chn(mess)=\\
\notag
&\{f\in Fresh \:|\: \exists s\in tmf.(h(s)\sqsubseteq_{acc}mess\wedge f\sqsubseteq_{acc}s)\}\\
\notag
&cs(claim,i)=\\
\notag
&\{f\in Fresh\;|\;cl=(i,secret,f)\wedge cl\in claim\}\\
\notag
&cc(claim, i, r)=\\
\notag &\{f\in Fresh\;|\; cl=(i,commit,r,f)\wedge cl\in claim \}
\end{align}

For $f\in Fresh$, $sk$, $pk$ denote the initiator's secret key and
responder's public key, $ps$ represents that the fresh was encrypt
by public key first and then secret key, and it is similar with
$sp$. We define $fen$ function as encryption type of some fresh f in
message.

\par\noindent \\$fen(f,mess)=$
\begin{equation}
\notag \left\{
\begin{aligned}
&sk,&\exists s\in tms.(f&\sqsubseteq_{acc} s\wedge s\sqsubseteq_{acc}mess),\\
&pk,&\exists s\in tmp.(f&\sqsubseteq_{acc} s\wedge s\sqsubseteq_{acc}mess),\\
&ps,&\exists s\in tmps.(f&\sqsubseteq_{acc} s\wedge s\sqsubseteq_{acc}mess),\\
&sp,&\exists s\in tmsp.(f&\sqsubseteq_{acc} s\wedge
s\sqsubseteq_{acc}mess).\\
\notag
&NULL,&otherwise
\end{aligned}
\right.
\end{equation}

Then we define $enc$ to encrypt $f$ with $s$ in messages. If $f$ has
been encrypt by $s$ already, then do nothing.

\par \noindent
\\$enc(f,s,mess)=$
\begin{equation}
\notag \left\{
\begin{aligned}
&mess,&fen(f,mess)=s\; or\; ps\; or\; sp,\\
&mess[f/\{f\}_s],&otherwise.
\end{aligned}
\right.
\end{equation}

We define $eha$ to encrypt fresh set $F$ with hash
function. Let $F \subseteq Fresh$.

\par \noindent
\\$eha(F,mess)=$
\begin{equation}
\notag \left\{
\begin{aligned}
&mess,&F\setminus \{f\in F|f\in chn(mess)\}=\emptyset,\\
&mess\cdot h(F),&otherwise\\
\end{aligned}
\right.
\end{equation}

\subsubsection{Algorithms}

According to the transformation techniques presented in Section 4,
we have designed algorithms for enhancing the security level of
protocols. The pseudo-codes of the algorithms are in the next page.
In the algorithms, $i$ denotes the initiator and $r$ the responder.

\bigskip
\par \noindent \textbf{Algorithm 1}
This algorithm is based on Proposition 1 for ensuring secrecy under
AKC. %
The algorithm works as follows: we set $secret\_set\_ini$ and $secret\_set\_res$ to store freshes claims to be secret in initiator and
responder. We go through each message, encrypt fresh in $secret\_set\_ini$ or $secret\_set\_res$ with secret short-term key which is generated by the other opposite party.

\begin{algorithm}[htb]
\caption{ transform-two-party-secrecy (protocol)}
\label{alg:Framwork}
\begin{algorithmic}[1]
\STATE $secret\_set\_ini=cs(claim,i)$
\STATE $secret\_set\_res=cs(claim,r)$
\IF{$secret\_set\_ini\neq\emptyset$\;or\;$secret\_set\_res\neq\emptyset$}
\FOR{each $m\in mess$} \STATE $ft=cn(m)$
\IF{m is the first message}
\STATE $m_a=i,r,Request$
\STATE
$m_b=r,i,\{ni\}_{pk(i)}$
\STATE Insert two messages
$m_a$, $m_b$ before m
\FOR {each $n\in secret\_set$}
\STATE $enc(n,ni,m)$
\ENDFOR
\ELSE
\IF{m is
transmitted from i to r}
\STATE k is a secret short-term key generated by r
\STATE $secret\_set=secret\_set\_ini$
\ELSE
\STATE k is a secret short-term key generated by i
\STATE $secret\_set=secret\_set\_res$
\ENDIF
\FOR {each $n\in secret\_set$}
\STATE $enc(n,k,m)$
\ENDFOR
\ENDIF
\ENDFOR \ENDIF
\end{algorithmic}
\end{algorithm}

\bigskip
\par \noindent \textbf{Algorithm 2}
This algorithm is based on Proposition 2 for ensuring the
commit-property. The algorithm also go through each message, and
encrypt fresh with secret key
or hash function. We set $com\_set\_ini$ and $com\_set\_res$ to store freshes claims to commit in initiator and
responder and assume secret values $ni$ and $nr$. If the fresh is encrypted by secret key, then algorithm will follow Proposition 11. Otherwise, it will follow Proposition 2.

\begin{algorithm}[htb]
\caption{ transform-two-party-commit (protocol)}
\label{alg:Framwork}
\begin{algorithmic}[1]
\STATE $com\_set\_ini=cc(claim,i,r)$
\STATE $com\_set\_res=cc(claim,r,i)$
\STATE $ni$ is a secret short-term
key for i
\STATE $nr$ is a secret short-term key for r
\IF {$com\_set\_ini\neq\emptyset$ or
$com\_set\_res\neq\emptyset$}
\FOR{each $m\in mess$}
\STATE
$ft=cn(m)$
\IF{m is transmitted from i to r}
\STATE $com\_set=com\_set\_res$
\STATE $ns=ni$
\ELSE
\STATE
$com\_set=com\_set\_ini$
\STATE $ns=nr$ \ENDIF
\FOR{each $n\in ft$} \IF{$n\in com\_set$}
\IF{$fen(n,m)=NULL$} \STATE $enc(n,sk,m)$ \ELSE
\IF{$fen(n,m)=pk$} \STATE $eha(\{ns,n\},m)$ \ENDIF
\ENDIF \ENDIF \ENDFOR \ENDFOR \ENDIF
\end{algorithmic}
\end{algorithm}


\begin{thebibliography}{X}
\bibitem{01} Blake-Wilson, Simon, D. Johnson, and A. Menezes. "Key agreement protocols and their security analysis." Lecture Notes in Computer Science (1997):30-45.
\bibitem{02} David Basin, Cas Cremers, and Marko Horvat. "Actor Key Compromise: Consequences and Countermeasures." 2014 IEEE 27th Computer Security Foundations Symposium (CSF) IEEE Computer Society, 2014:244-258.
\bibitem{03} Gao, Meng, and F. Zhang. "Key-Compromise Impersonation Attacks on Some Certificateless Key Agreement Protocols and Two Improved Protocols." Education Technology and Computer Science, International Workshop on IEEE, 2009:62-66.
\bibitem{04} Qiang Tang, and Liqun Chen. "Extended KCI attack against two-party key establishment protocols." Information Processing Letters 111.15(2011):744¨C747.
\bibitem{05} Chalkias, K., et al. "Two Types of Key-Compromise Impersonation Attacks against One-Pass Key Establishment Protocols." e-Business and Telecommunications e-Business and Telecommunications, 2009:227.
\bibitem{06} Chalkias, K., et al. "On the Key-Compromise Impersonation Vulnerability of One-Pass Key Establishment Protocols" 2007 SECRYPT, 2007:222-228
\bibitem{07} Gorantla, M. C., et al. "Modeling key compromise impersonation attacks on group key exchange protocols." Lecture Notes in Computer Science 14.4(2009):105-123.
\bibitem{08} Lamacchia, B., K. Lauter, and A. Mityagin. "Stronger security of authenticated key exchange¡¯, Paper presented." The Proceedings of The Provsec¡¯07 of Lncs 2006.1-4(2007):263-283(21).
\bibitem{09} Zhu, Robert W., Tian, Xiaojian and Wong, Duncan S.. "Enhancing CK-Model for Key Compromise Impersonation Resilience and Identity-based Key Exchange.." IACR Cryptology ePrint Archive 2005 (2005): 455.
\bibitem{10} Basin, David, and C. Cremers. "Modeling and Analyzing Security in the Presence of Compromising Adversaries." Lecture Notes in Computer Science (2010).
\bibitem{11} Cremers, Cas, and S. Mauw. Operational semantics and verification of security protocols. Operational semantics and verification of security protocols. Springer, 2012.
\bibitem{12} Shim, Kyungah. The Risks of Compromising Secret Information. Information and Communications Security. Springer Berlin Heidelberg, 2002.
\bibitem{13} Chalkias, K., et al. "Two Types of Key-Compromise Impersonation Attacks against One-Pass Key Establishment Protocols." e-Business and Telecommunications e-Business and Telecommunications, 2009:227.
\bibitem{14} Cremers, C., Mauw, S. "A Family of Multi-Party Authentication Protocols".
First Benelux Workshop on Information and System Security (WISSec) (2006)
\bibitem{15} Basin, David, et al. "Improving the Security of Cryptographic Protocol Standards." Security and Privacy, IEEE 13(2015).
\bibitem{16} Aaron D. Jaggard, et al. "Breaking and fixing public-key Kerberos." In Proc. WITS¡¯06 2006:402¨C424.
\bibitem{17} Bella, Giampaolo, F. Massacci, and L. C. Paulson. "An overview of the verification of SET." International Journal of Information Security 4.1-2(2005):2005.
\bibitem{Scyther} C. Cremers, ¡°The Scyther Tool: Verification, falsification, and analysis
\bibitem{19} "Security Protocols Open Repository", http://www.lsv.ens-cachan.fr/Software/spore/index.html
of security protocols,¡± in Proc. CAV, ser. LNCS, vol. 5123.
Springer, 2008, pp. 414¨C418.
\bibitem{20}Basin, David, C. Cremers, and S. Meier. "Provably repairing the ISO/IEC 9798 standard for entity authentication." Proceedings of the First international conference on Principles of Security and Trust Springer-Verlag, 2012:129-148.
\report
\end{thebibliography}
\end{document}